\documentclass{aa}  

\usepackage{hyperref}
\usepackage{graphicx}
\usepackage{caption}
\usepackage{tabularx}
\usepackage{xcolor}
\usepackage{txfonts}
\usepackage{xspace}

\def\HI{\ion{H}{I}}

\newcommand{\CO}{\element[]CO}
\newcommand{\couno}{\element[][12]CO (1--0)\,}

\def\halpha{\ion{H$\alpha$}}
\def\hbeta{\ion{H$\beta$}}
\def\nad{\ion{NaI}~D}
\def\OIst{[\ion{O}{I}]$\lambda$6300}
\def\OIsts{[\ion{O}{I}]$\lambda$6364}
\def\OIIIqn{[\ion{O}{III}]$\lambda$4959}
\def\OIIIfs{[\ion{O}{III}]$\lambda$5007}
\def\NIIscq{[\ion{N}{II}]$\lambda$6548}
\def\NIIsco{[\ion{N}{II}]$\lambda$6583}
\def\SIIuno{[\ion{S}{II}]$\lambda$6716}
\def\SIIdue{[\ion{S}{II}]$\lambda$6731}
\def\SIIdoublet{[\ion{S}{II}]$\lambda\lambda$6716,6731}

\def\OI{[\ion{O}{I}]}

\def\OIII{[\ion{O}{III}]}
\def\NII{[\ion{N}{II}]}
\def\SII{[\ion{S}{II}]}

\usepackage{printlen}
\usepackage{layouts}

\def\NII{[N{\,\small II}]}

\def\Htwo{H$_{\,2}$}

\newcommand{\kms}{$\,$km~s$^{-1}$}

\newcommand{\Jyb}{Jy\,beam$^{-1}$}

\newcommand{\mJyb}{mJy\,beam$^{-1}$}

\newcommand{\msun}{{${\rm M}_\odot$}}
\newcommand{\msunyr}{{${\rm M}_\odot$ yr$^{-1}$}}

\newcommand{\cmsq}{cm$^{-2}$}
\newcommand{\pcdue}{pc$^{-2}$}

\newcommand{\eg}{\mbox{e.g.}}
\newcommand{\ie}{\mbox{i.e.}}

\newcommand{\forn}{\mbox Fornax~A}

\newcommand{\meer}{{MeerKAT}}

\newcommand{\vsys}{{$v_{\rm sys}$}\,}

\begin{document} 

   \title{AGN feeding and feedback in Fornax~A:}
   \titlerunning{Feeding and feedback in Fornax~A:}
   \subtitle{Kinematical analysis of the multi-phase ISM}

        \author{F. M. Maccagni\inst{1},
        P. Serra\inst{1},
        M. Gaspari\inst{2,3},
        D. Kleiner\inst{1},
        K. Morokuma-Matsui\inst{4},
        T. A. Oosterloo\inst{5,6},
        M. Onodera\inst{7,8},
        P. Kamphuis\inst{9},
        F. Loi\inst{1},
        K. Thorat\inst{12},
        M. Ramatsoku,\inst{10,1},
        O. Smirnov\inst{10,11},
        S. V. White\inst{10}}

   \institute{INAF -- Osservatorio Astronomico di Cagliari, via della Scienza 5, 09047, Selargius (CA), Italy
         \and
         INAF -- Osservatorio di Astrofisica e Scienza dello Spazio di Bologna, via Piero Gobetti 93/3, I-40129 Bologna, Italia
         \and
         Department of Astrophysical Sciences, Princeton University, 4 Ivy Lane, Princeton, NJ 08544-1001, USA
         \and
            Institute of Astronomy, Graduate School of Science, The University of Tokyo, 2-21-1 Osawa, Mitaka, Tokyo 181-0015, Japan        
         \and
  			Netherlands Institute for Radio Astronomy (ASTRON), Oude Hoogeveensedijk 4, 7991 PD, Dwingeloo, The Netherlands
  		 \and
  			Kapteyn Astronomical Institute, University of Groningen, Landleven 12, NL-9747 AD, Groningen, the Netherlands
   		 \and
  			Subaru Telescope, National Astronomical Observatory of Japan, National Institutes of Natural Sciences (NINS), 650 North A'ohoku Place, Hilo, HI 96720, USA
  		  \and
 			Department of Astronomical Science, The Graduate University for Advanced Studies, SOKENDAI, 2-21-1 Osawa, Mitaka, Tokyo, 181-8588, Japan
 		 \and
 		    Ruhr University Bochum, Faculty of Physics and Astronomy, Astronomical Institute, 44780 Bochum, Germany
 		  \and
 		  Department of Physics, University of Pretoria, Private Bag X20, Hatfield 0028, South Africa
 		  \and
 		  Department of Physics and Electronics, Rhodes University, PO Box 94, Makhanda, 6140, South Africa
 		  \and
 		  South African Radio Astronomy Observatory, 2 Fir Street, Black River Park, Observatory, Cape Town, 7925, South Africa
    \\
   \email{filippo.maccagni@inaf.it}
             }

   \date{Received April 21, 2021 accepted August 11, 2021}

     \abstract
   {We present a multi-wavelength study of the gaseous medium surrounding the nearby 
   active galactic nucleus (AGN) \forn. Using MeerKAT, ALMA and MUSE observations we reveal a complex distribution of the atomic (\HI), molecular (\CO), and ionised gas in its centre and along the radio jets. By studying the multi-scale kinematics of the multi-phase gas, we reveal the presence of concurrent AGN feeding and feedback phenomena. Several clouds and an extended 3\,kpc filament -- perpendicular to the radio jets and the inner disk ($r\lesssim 4.5$ kpc) -- show highly-turbulent kinematics, which likely induces nonlinear condensation and subsequent Chaotic Cold Accretion (CCA) onto the AGN. In the wake of the radio jets and in an external ($r\gtrsim 4.5$ kpc) ring, we identify an entrained massive ($\sim$\,$10^7$\ \msun) multi-phase outflow ($v_{\rm OUT}\sim 2000$~\kms). The rapid flickering of the nuclear activity of \forn~($\sim$\,3 Myr) and the gas experiencing turbulent condensation raining onto the AGN provide quantitative evidence that a recurrent, tight feeding and feedback cycle may be self-regulating the activity of \forn, in agreement with CCA simulations. To date, this is one of the most in-depth probes of such a mechanism, paving the way to apply these precise diagnostics to a larger sample of nearby AGN hosts and their multi-phase ISM. 
   } 

	\titlerunning{Feeding and Feedback in \forn}
	\authorrunning{F. M. Maccagni, P. Serra, M. Gaspari, et al.}
   \maketitle

\section{Introduction}
\label{sec:intro}

Feedback from active galactic nuclei (AGNs) is one of the key processes that can affect the evolution of a galaxy. The energy released by the supermassive black hole (SMBH) as radiation and/or jets of radio plasma induces disturbances in the hot halo of its host galaxy, generating cavities and shocks which prevent the hot gas from cooling~\citep[][]{Fabian:2012,McNamara:2012,Baldi:2019}. The energy of AGNs is sufficient to quench star formation and cooling flows in massive galaxies and groups/clusters~\citep[][]{Churazov:2012,Cano-Diaz:2012,Carniani:2016,McDonald:2018}, mainly through the ejection of jets and circulation of massive gaseous outflows~\citep[see, for example, ][for different reviews on the topic]{King:2015_rev,Harrison:2017,Morganti:2018,Veilleux:2020}.

In some models of galaxy evolution, multiple feedback episodes are needed to prevent cooling from the hot halo and efficiently regulate star-formation~\citep[see, for example, ][]{Ciotti:2010,Gaspari:2017_cca,Prieto:2021}. Since AGNs are associated with the accretion of material onto the SMBH, recurrent feedback is generated by recurrent accretion events (\ie\ recursive feeding).
The existence of recurrent AGNs has long been known~\citep[\eg][]{Cordey:1987,Saikia:2009,Shulevski:2017,Brocksopp:2007,Orru:2015}. Recently, low-frequency radio surveys have shown that multiple phases of activity occur in a non-negligible number of sources~\citep[][]{Sabater:2019,Morganti:2020} and short timescales of activity, between 1 and 100 Myr, were derived~\citep[][]{Brienza:2017,Sabater:2019}. 

\begin{figure}
	\centering
	\includegraphics[trim = 0 0 0 0, width=\columnwidth]{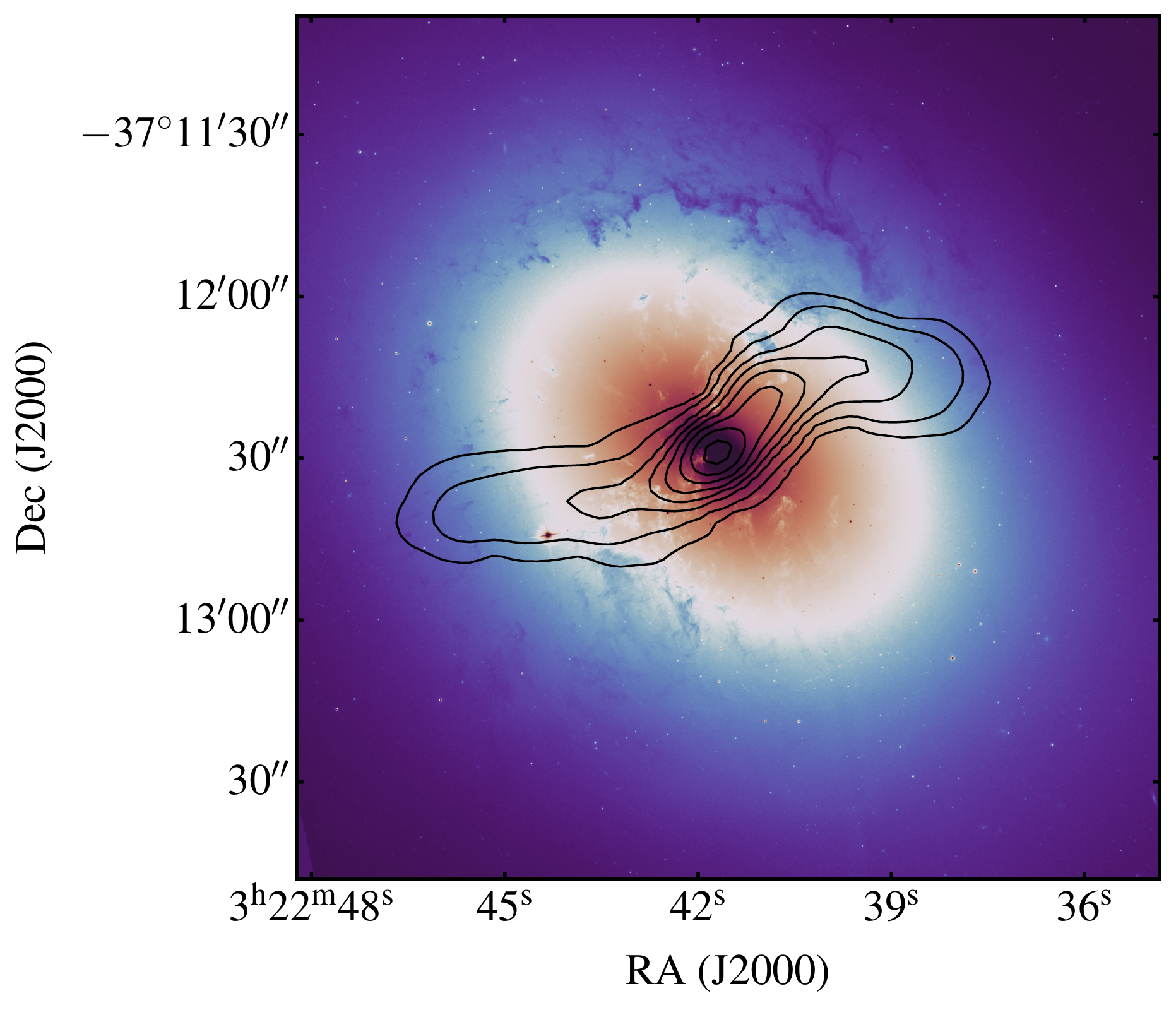}
	\caption{Hubble ACS visible emission (535 nm) of the centre of \forn, radio jets seen by \meer\ at 1.4 GHz are overlaid in black. Contour levels are $5 \times 10^{-4}\times 2^n$\,\Jyb\ (n $= 0, 1, 2,$...)~\citep{Maccagni:2020}.}
	\label{fig:HST}
\end{figure}

Multi-wavelength spectral observations are essential for a complete characterisation of feeding and feedback phenomena. High-resolution observations of the kinematics of the gas are crucial to distinguish between feeding (\ie\ in-flows of gas), and feedback processes (\ie\ gaseous outflows).

Gaseous outflows are known to be multi-phased, from the hot, highly-ionised phase \citep[T$\gtrsim 10^{6-8}$K; \eg][]{Tombesi:2013}, to the warm ionised~\citep[T$\sim 10^{3-6}$K; \eg][]{Mingozzi:2019,Davies:2020}, neutral atomic~\citep[T$\sim 10^{2-3}$K; \eg][]{Morganti:2018,Roberts-Borsani:2019} and molecular phase~\citep[T$\lesssim 10^{2}$K eg][]{Fluetsch:2020,Veilleux:2020}. A tight correlation between the distribution and kinematics of these gaseous phases has been observed in several galaxies~\citep[see, for example,~\mbox{NGC 1266},~\mbox{Mrk 231},~\mbox{IC 5063} in][respectively]{Alatalo:2011,Veilleux:2016,Oosterloo:2019}, with the cold neutral gas seemingly the most massive~\citep[][]{Cicone:2018,Fluetsch:2020,Veilleux:2020}.
 
In self-regulated feedback scenarios, outflows inject turbulence in the surrounding medium causing the hot gas to condense from the hot phase into cold atomic and then molecular clouds, which then `rain' back growing the central SMBH~\citep{Gaspari:2019}. In a handful of galaxies, both feeding and feedback processes have been associated with the same episode of activity~\citep[see, for example NGC 1433, NGC 1068 and NGC1808 in][respectively]{Combes:2013,Garcia-Burillo:2019,Audibert:2020}, but a connection with a previous (or future) activity and the sustainability of recursive triggering has not been made.

Chaotic Cold Accretion (CCA; \citealt{Gaspari:2013_cca,King:2015,Gaspari:2015_cca,Gaspari:2017_cca,Prasad:2017}) is one of the few major feeding mechanisms that can generate rapid recursive nuclear activity, self-consistently. In this process, cold clouds and filaments condense out of the hot phase via turbulent nonlinear thermal instability. Chaotic collisions promote the `funneling' of the cold phase toward the SMBH, and ignite the AGN~\citep{Gaspari:2012b,Yang:2019}. Feedback from the AGN outflows re-heat the core, while entraining the ambient gas, establishing a tight AGN self-regulated loop~\citep{Gaspari:2017_uni}. Hence, linking chaotic accretion to the cooling rate induces a natural self-regulation, with a flickering recursive cycle and peaks of AGN activity.

The CCA model provides key predictions for the kinematics of all phases of the gas, and the correlations between them, which can be directly compared with the results of multi-wavelength spectral observations~\citep{Gaspari:2018}. There is multiple evidence of cooling and inflows connected to CCA in the kinematics of several phases of the interstellar, intragroup and intracluster medium~\citep[\HI, \CO, H$\alpha$ gas, and X-ray halos;][]{Rose:2019,Storchi-Bergmann:2019} and in different types of host galaxies, from isolated early-type galaxies (\eg\ \citealt{Maccagni:2014,Maccagni:2016,Maccagni:2018,Lakhchaura:2018}) to group galaxies (see, for example, \citealt{Temi:2018,Juranova:2019,Juranova:2020}) and brightest cluster galaxies (\eg\ \citealt{Tremblay:2016,Tremblay:2018,Olivares:2019}). However, it is challenging to trace this phenomenon from the macro-scale (tens of kpc) to the micro-scale (sub-pc) where the accretion onto the SMBH occurs~\citep[see][for a review]{Gaspari:2020}. So far, most detections trace in-falling gas in the innermost 100 pc, or average properties of the $\sim$\,1-20 kpc halo have been used to infer the presence of cooling caused by turbulence.

Recurrent nearby AGNs are ideal to study in detail feeding and feedback mechanisms, quantify their impact on the evolution of the galaxy and understand how they can self-regulate their recursive activity. Among these AGNs, the nearby \citep[$D_{\rm L}=20.8$ Mpc][]{Cantiello:2013} radio galaxy \forn\ is a pivotal target.
The radio spectral distribution of \forn\ shows indications of rapid flickering~\citep[][]{Maccagni:2020}. Its giant radio lobes suggest they formed through multiple episodes, the latter of which was interrupted 200 Myr ago. More recently ($\sim 3$ Myr ago), a less powerful and short ($1$ Myr) phase of nuclear activity generated the central s-shaped jets, which are now expanding in the Inter-Stellar Medium (ISM). Currently, the core may be in a new active phase. 
\begin{table*}[tbh]
        \caption{Main properties of the observations of the multi-phase gas in \mbox{NGC~1316}.}
        \centering
        \label{tab:forObs}
        \begin{tabularx}{\textwidth}{X X X c c c}  
                \hline\hline                                                         
                Telescope  			& Observing time	& Bandwidth     & Spatial resolution			&  Spectral resolution 		\\
                \hline
                \meer\	    			& 	8+7	hrs			&	   1.402 - 1.420 GHz    	&$22$\arcsec$\times18$\arcsec   &     	209 kHz ($\sim 44$~\kms)		        			\\
                ALMA		&	5.7	hrs			&		112.8 - 114.7 GHz 		&$22$\arcsec$\times18$\arcsec   &				        3.8 kHz ($\sim 9.9$~\kms)	\\  
                MUSE  wide-field$^\star$	    &   333s			&4650--9300~\AA	&$2.1\arcsec$ (seeing)	        &$2.3$~\AA\ ($\sim 68$~\kms)&	\\             
                MUSE  centre-field$^\star$	  &   2178s			&4650--9300~\AA	&$2.3\arcsec$ (seeing)	        &$2.3$~\AA\ ($\sim 68$~\kms)&	\\
                \hline                           
        \end{tabularx}
         \tablefoot{(*) surface brightness limit after Galactic extinction correction for \hbeta\ and \halpha\ lines. The MUSE spectral resolution ($\sigma$) is given at the \NIIsco\ line.}
\end{table*}

A major merger occurred $1$--$3$ Gyr ago~\citep[\eg][]{Schweizer:1980,Serra:2019}. This event likely caused the first triggering of the radio source and brought large amounts of dust~\citep[][]{Schweizer:1980,Lanz:2010,Galametz:2012,DuahAsabere:2016}, molecular gas~\citep[][]{Horellou:2001,Galametz:2014,Morokuma:2019}, and neutral hydrogen~\citep{Horellou:2001,Serra:2019,Richtler:2020,Kleiner:2021} into the centre and around the galaxy. Fig.~\ref{fig:HST} shows that the dust seen in extinction in the innermost $\approx 8$ kpc is distributed mainly in two structures: a stripe crossing the centre from the north-west to the south-east; and two outer arcs that seem to originate at the bend of the radio jets. \halpha, \CO\ and \HI\ are detected both on the stripe and along the outer arcs~\citep[][]{Mackie:1998,Morokuma:2019,Serra:2019}. The molecular gas shows clouds with irregular kinematics on both features, suggesting a tight interplay between the expansion of the radio jets and the surrounding interstellar medium~\citep[][]{Morokuma:2019}. 

In this paper, we analyse in detail the kinematics of the multi-phase gas in \mbox{NGC~1316} (\forn). We aim to understand if and how the cold and ionised gas have triggered the nuclear activities in the last $3$ Myr, and what are the feedback effects resulting from these. Section~\ref{sec:obsDR} describes the observations from state-of-the-art radio, millimetre and optical telescopes to study at high spatial and spectral resolution the multi-phase gas in \forn. Sects.~\ref{sec:momsCold} and~\ref{sec:momsIon} show the distribution, velocity field and dispersion maps of the \HI, CO and ionised gas detected in the innermost arcminute of the galaxy. The detailed analysis of their kinematics is discussed in Sect.~\ref{sec:kinIon}. In Sect.~\ref{sec:discussion}, we reflect on how the AGN and its jets affect the gas kinematics, and viceversa, how the gas may fuel the AGN and influence the expansion of the radio jets revealing the presence of both feeding and feedback phenomena. Sect.~\ref{sec:conclusions} summarises the most important results of this work.

\section{Observations and data reduction}
\label{sec:obsDR}

For the purposes of this paper we consider high resolution spectral observations from the radio, millimetre and optical wavelengths. We use \meer\ observations to study the neutral hydrogen (\HI) 21-cm line, the Atacama Large Millimeter and sub-millimeter Array (ALMA) for the \couno\ emission and the Multi Unit Spectroscopic Explorer (MUSE) on the Very Large Telescope (VLT) for the ionised lines (Balmer lines, \OIII, \OI, \NII\ and \SII).

Even though these instruments are the ones producing the highest spectral and spatial resolutions in their respective bands, we point out that the spatial resolution of the \meer\ and ALMA observations ($\sim 20$\arcsec) is 10 times lower than the one of MUSE ($\sim 2$\arcsec) and their spectral resolution is between 2 and 5 times better. On the other hand, the \meer\ \HI\ and ALMA CO observations have a wide field of view (FoV, 1$^\circ$ and $8$\arcmin, respectively), while MUSE observations are limited to the innermost arcminute of AGNs. Because of the proximity of \forn, the combination of all three observations allows us to study in detail the ISM from its innermost kilo-parsec to the periphery of its stellar body. In Table~\ref{tab:forObs}, we summarise the main properties of the \meer, ALMA and MUSE observations.

\subsection{Neutral atomic hydrogen observations}
\label{sec:hiObs}
We observed \forn\ with \meer\ \citep{Jonas:2016,Camilo:2018} in two different commissioning observations in June 2018. The observations were carried out with a different number of antennas (36 and 64, respectively) using the 4096-channels correlator set in the frequency interval 856–1712 MHz in full polarisation. The channel width is 209 kHz, corresponding to 44.5~\kms\ for \HI\ at redshift z = 0. The total integration time on \forn\ was 8+7 hours. 

 A complete description of the data reduction of the 21-cm observations can be found in \citet[][]{Kleiner:2021}. The data reduction was made with {\tt CARACal}\footnote{\url{https://github.com/caracal-pipeline/caracal}}~\citep[][]{Jozsa:2020}. {\tt CARACal} is a containerised pipeline written in {\tt Python3} and has been used to reduce several radio continuum and spectral observations from MeerKAT~\citep[\eg][]{Ramatsoku:2020a,deBlok:2020,Kleiner:2021} the Very Large Array~\citep[][]{Ramatsoku:2020b} and the upgraded Giant Meterwave Radio Telescope~\citep[][]{Michalowski:2019}. 

 For the purposes of this paper, we generated a datacube using a smaller arcsecond-tapering than the one used in \citet[][]{Kleiner:2021}, which allows us to obtain a higher spatial resolution of $22\arcsec\times18\arcsec$. The noise in the cube is $0.1$~\mJyb\ per channel. This corresponds to a $3\sigma$~\HI\ column density sensitivity of $3.2\times 10^{19}$~\cmsq\ in a single channel, and to a surface brightness sensitivity limit of $0.25$~\msun\,\pcdue. Assuming a $100$\kms\ line-width, the $5\sigma$ point-source $M_{\rm HI}$ sensitivity is $3.0\times 10^6$~\msun\ at the assumed distance of 20.8 Mpc.

\begin{figure*}
	\centering
	\includegraphics[]{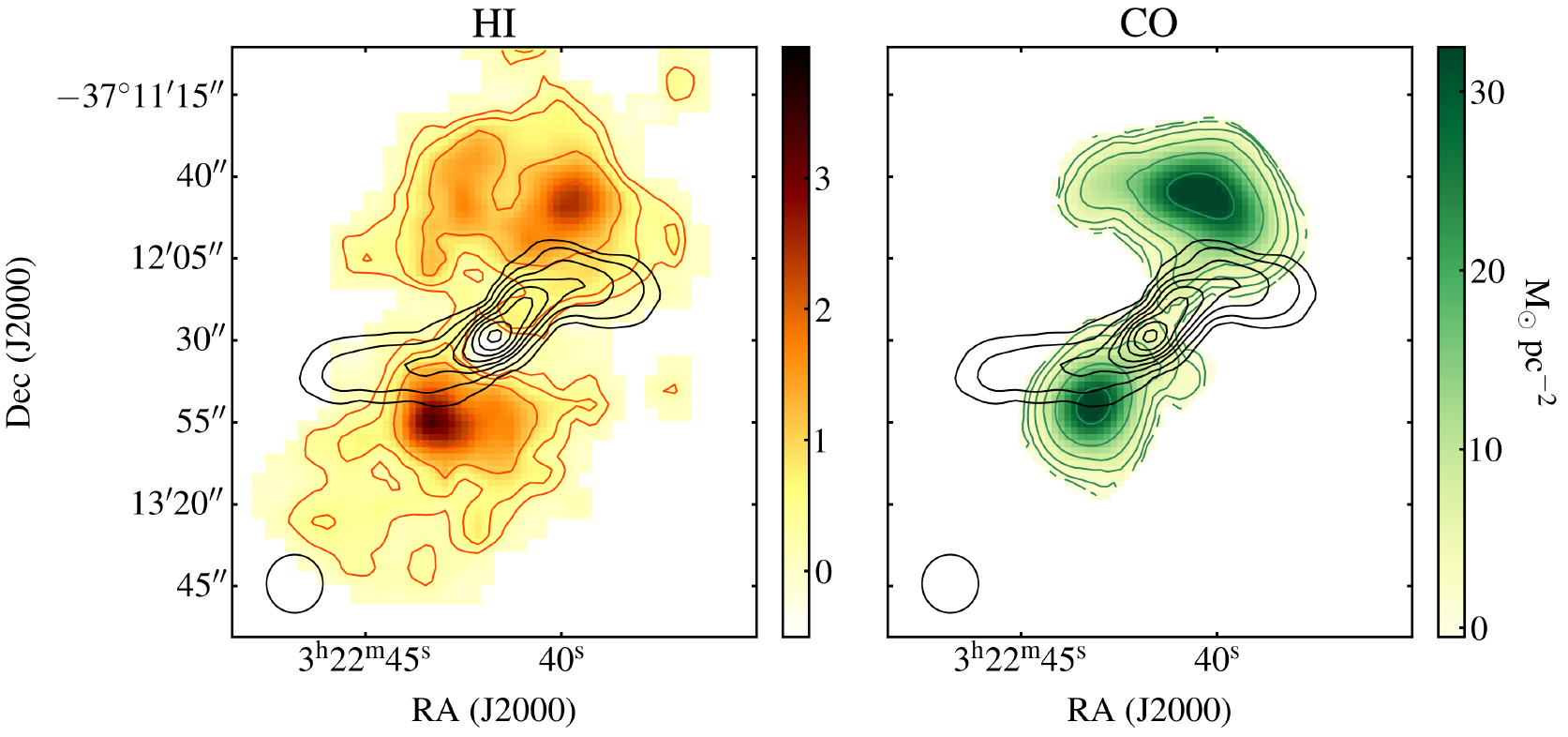}
	\caption{ The {\em left panel} shows the surface brightness map of the \HI\ gas in the centre of \forn. The {\em right panel} show the same maps for the \couno. The PSF of all maps is at $22\arcsec\times18\arcsec$. Surface brightness contour levels are $3\sigma \times 2^n$\,\msun\,\pcdue\ (n $= 0, 1, 2,$...) in all panels. The lowest contour corresponds to the $3\sigma$ surface brightness detection limit. Radio jets are the same as in Fig.~\ref{fig:HST}.}
	\label{fig:momsCold}
\end{figure*}

 Since we knew of the presence of \HI\ in the centre of \forn\ from previous \meer\ observations~\citep{Serra:2019}, we fine tuned the parameters of the source-finding algorithm {\tt SoFiA}\footnote{\url{https://github.com/SoFiA-Admin}}~\citep[][]{Serra:2015} to retrieve all the already-known \HI\ and investigate the significant presence of a more diffuse component previously undetected. The resulting surface brightness map, velocity field and velocity-dispersion field are shown in Fig.~\ref{fig:momsCold}. 

\subsection{Molecular gas observations}
\label{sec:coObs}

A complete description of the reduction of the ALMA 108 GHz observation can be found in~\citet{Morokuma:2019}. The PSF of the resulting datacube is very asymmetric ($15.5$\arcsec$\times7$\arcsec), hence, for a quantitative comparison between the atomic and molecular phase, we generate a new \couno\ datacube at the same spatial resolution of the \HI. We create the total \couno\ image, velocity field and velocity-dispersion field using {\tt SoFiA}, analogously to what done for the \HI. We convert the brightness temperature map ($T_{\rm B,CO}$) in surface brightness units assuming a Galactic  \couno-to-\Htwo\ conversion factor $X_2=4.3$\msun (K~\kms\ \pcdue)$^{-1}$. The noise in the datacube is $12$~\mJyb\ per channel ($\Delta v\sim9.9$\kms). This corresponds to a $3\sigma$ surface brightness limit of $0.33$~\msun\,\pcdue.

\begin{figure*}
	\centering
	\includegraphics[]{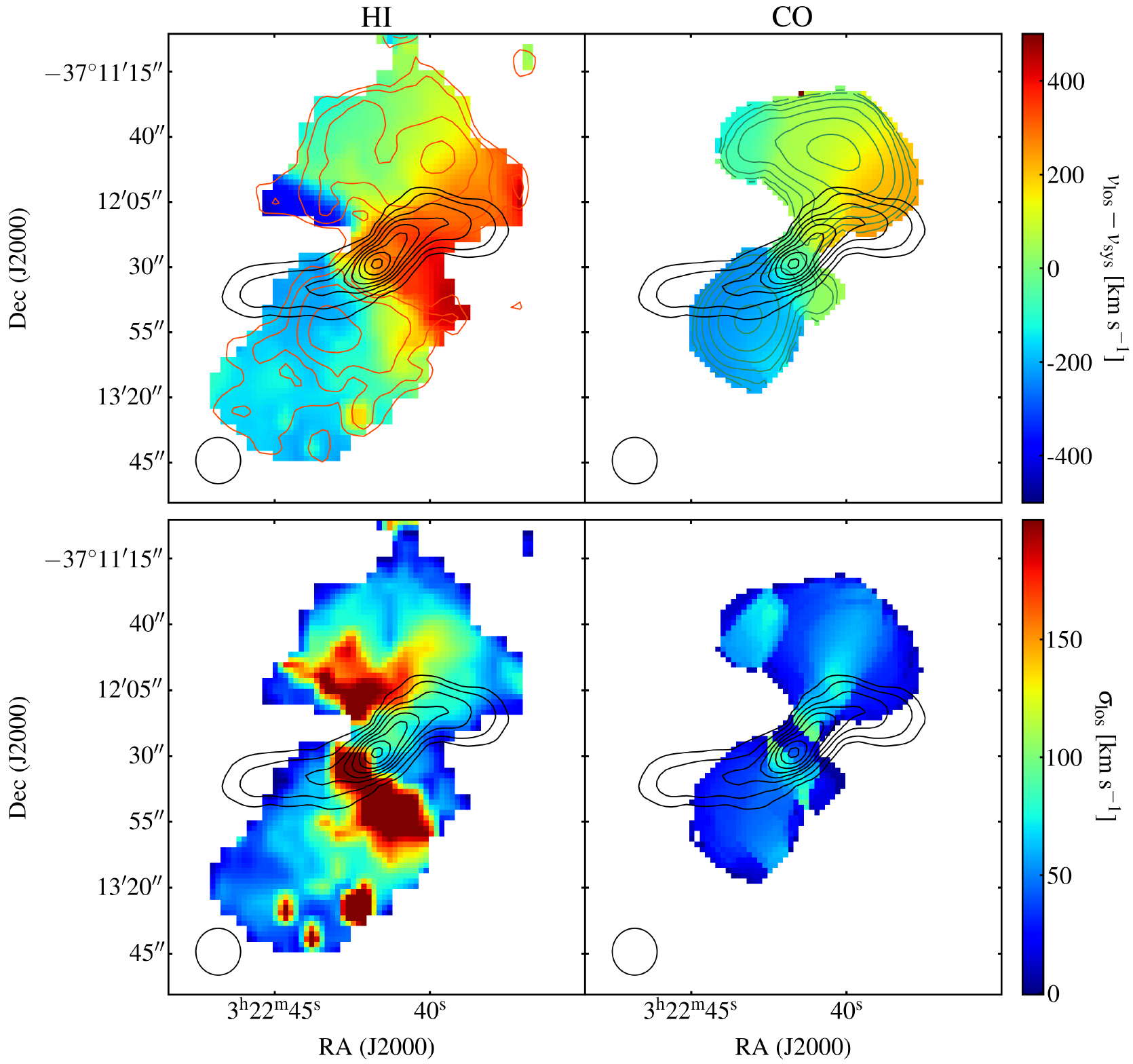}
	\caption{The {\em left panels} show the velocity and the velocity dispersion fields of the \HI\ gas in the centre of \forn. The {\em right panels} show the same maps for the \couno. The PSF of all maps is $22\arcsec\times18\arcsec$. Surface brightness contour levels and radio jets are as in Fig.~\ref{fig:momsCold}.}
	\label{fig:momsCold12}
\end{figure*}

\subsection{Ionised gas observations}
\label{sec:ionGasObs}

To study the ionised gas in \forn, we select two different sets of MUSE observations available on the ESO archive\footnote{\url{http://archive.eso.org/scienceportal/home}}. The wavelength coverage of the MUSE cubes ranges from 4650~\AA\ to 9300~\AA. Within this interval MUSE achieves a resolution of $R\sim 1750$--$3750$, which at the \NIIsco\ line corresponds to approximately $68$~\kms.

The first dataset consists of four different observations of the full stellar body of \forn. These observations were taken on November 11$^{\rm th}$ 2014 and December $13^{\rm th}$ 2014 (proj. ID: 094.B-0298, PI Walker, C.J.). The total integration time per observation is $333$s. The spatial resolution of the observations given by the seeing is $\approx 2.1$\arcsec. 

In all datacubes, we detect the Balmer lines (\halpha, \hbeta) as well as the forbidden transitions of the ionised metals \OIIIqn, \OIIIfs, \OIst, \OIsts, \NIIscq, \NIIsco, \SIIdoublet\ and the \nad\ absorption doublet of neutral gas. As available on the archive, the 4 mosaics of the single MUSE pointings have strong sky variations throughout the field of view. The low signal-to-noise of most detected lines does not allow a complete characterisation of their total flux and kinematical properties (\ie\ line-shift and line-width). For this reason, we use this mosaic only to describe the overall kinematics of the ionised gas throughout the entire stellar body by focusing only on the \NIIsco\ line (see Sect.~\ref{sec:kinIon}).

To estimate and subtract the stellar continuum, the datacubes are binned using Voronoi tessellation to achieve an average signal-to-noise ratio (S/N) of 30 per wavelength channel in each bin between 4650--6800~\AA. In this step, all spectra of the datacube with S/N$\lesssim$3 are discarded. This step is performed using the {\tt GISTpipeline}\footnote{\url{http://ascl.net/1907.025}}~\citep[][]{Bittner:2019}. The pipeline models the stellar continuum using the MILES stellar library templates~\citep[][]{Sanchez-Blazquez:2006} which covers the wavelength range from 3525 -- 7500\AA. The fitting is performed using an adapted Penalized Pixel-Fitting routine~\citep[pPFX, ][]{Cappellari:2004,Cappellari:2017} and allowing an additive Legendre polynomial of $11^{\rm th}$ order to correct for the shape of the continuum template.

We mosaic together the four stellar-subtracted datacubes using~{\tt Montage}. From the mosaicked cube we extract a sub-cube centred on the \NIIsco\ line in the velocity range [-2100,+800]~\kms\ with respect to the systemic velocity of \mbox{NGC~1316} (\vsys$=1720$~\kms). In the rest of the paper we refer to this datacube as the {\em wide-field} MUSE datacube (which FoV is approximately $17.24\times 12.53$ kpc).

Recently,~\cite{Richtler:2020} presented the same observations focusing on the \NIIsco\ line and the \nad\ absorption doublet. Our study presents both points of agreement and tension with the conclusions of these authors, as detailed in Sects.~\ref{sec:kinIon} and~\ref{sec:discussion}. 

The second dataset consists of a deeper single pointing MUSE observation in the innermost arcminute of \forn. This observation was taken on July, 24$^{\rm th}$, 2018 (proj. ID: 0101.D-0748, PI K. Hanindyo) in four different exposures. The total integration time was $2178$s over the spectral range 4750 -- 9350~\AA. The spatial resolution given by the seeing is $\sim 2.3\arcsec$. We use the MUSE datacube available in the ESO archive~\citep[that has been produced using the MUSE pipeline v1.4 with default parameters;][]{Weilbacher:2014}.

We subtract the stellar continuum following the same procedure as for the wide-field datacube. In this observation the \OIIIfs\ line is detected on average with S/N$\sim 5$, when present. This allows us to avoid performing Voronoi binning on the stellar-subtracted datacube to reach a uniform S/N over the entire field of view. Instead, we select only those pixels where the signal-to-noise ratio of the \OIIIfs\ is higher than $3.5$. In the rest of the paper we refer to this dataset as the {\em centre-field} datacube.

\subsection{Estimate of the line emission parameters}
\label{sec:gaussFit}

In the centre-field datacube we measure the main kinematical properties of the ionised gas by simultaneously fitting, on a pixel-by-pixel basis, one or two Gaussian components to the \hbeta, \OIIIfs, \OIIIqn, \NIIscq, \halpha, \NIIsco\ and \SIIdoublet\ lines (we exclude the low S/N \OI\ lines). In the simultaneous fit, each component is tied to have the same velocity centroid and dispersion in all lines, while the amplitudes are allowed to vary freely. The fits are performed using the Python library {\tt lmfit}~\citep[][]{Newville:2014}. {\tt lmfit} provides non-linear optimisation and is built on the Levenberg-Marquardt algorithm. The fitting routine, computation of the residuals, as well as all outputs presented in this paper make use of the {\tt GuFo}\footnote{\url{https://github.com/Fil8/GuFo}} suite we developed.

Across the field of view, the emission lines show different properties. In some regions, they are narrow and single-peaked, in others they are shallow and broad, while in the centre they clearly show the presence of a significant second peak (see Appendix~\ref{app:specMUSE} for a collection of spectra). In order to choose which is the best fit solution, we perform two different fitting runs. In the first run we fit a single Gaussian component, with dispersion that can vary between $1$ and $1000$~\kms\ and centroid that may range within $\pm 700$~\kms\ with respect to \vsys. In the second run, we fit two Gaussian components by letting the $\sigma$ parameter of both components vary between $1$ and $1000$~\kms.

In each run, for every line, we determine the residuals of the fits in the $\pm 1200$~\kms\ velocity interval centred on the systemic velocity of \forn. The best-fit solution minimises the residuals of the \OIIIfs\ line. One Gaussian component is sufficient to describe the emission lines in most of the field of view, except in the centre and in other regions along the nearly edge-on disk (see Sect.~\ref{sec:momsIon}). For each line of sight (with S/N$_{\rm [OIII]\lambda 5007}\gtrsim 3.5$), we measure the following properties of each component and of the total fitted lines: flux, centroid, dispersion and $w80$ (the width including $80\%$ of the flux).  

Overall, we perform the fit just described in $\approx 10^5$ sight-lines. To reduce computing time, we performed the fitting using the common {\tt multiprocess} libraries of {\tt Python 3}. 

\section{Distribution and kinematics of the cold gas}
\label{sec:momsCold}

In Fig.~\ref{fig:momsCold} we show the distribution of the neutral and molecular gas seen at the same resolution ($22$\arcsec$\times18$\arcsec), along with the radio jets of \forn. The cold gas extends out to 6.5 kpc in the north and 8.0 kpc in the south. Both the \HI\ and \couno\ follow the dust distribution, mainly oriented nearly edge-on along the NS direction, with arcs at larger radii ($r\gtrsim 4.5$ kpc). The most prominent difference between the \HI\ and the \CO\ is that diffuse neutral hydrogen is detected also perpendicularly to the dust (in the NE-SW direction), while molecular gas is not. The \HI\ also extends further out in the south of the field of view, but the sensitivity of ALMA in these regions is much lower than in the centre.

We use the relations shown in \citet{Meyer:2017} to convert the flux density map into \HI\ column density and surface brightness units. The total \HI\ mass in the centre of \forn\ is $\sim 6.7\times10^7$~\msun. This is about $30\%$ more than what was measured in~\cite{Serra:2019}, indicating that a significant component of \HI\ in \forn\ has low-column densities (between $1-3\times 10^{19}$\cmsq).

From the integrated \couno\ line we measure a total \Htwo\ $=5.8\times 10^{8}$~\msun~\citep[following the classical relations shown in, for example, ][]{Bolatto:2013}. The result is in agreement with what measured by~\citet[][]{Morokuma:2019}, and comparable to previous single dish observations~\citep[][]{Wiklind:1989,Sage:1993,Horellou:2001}

The velocity fields of the \HI\ and \CO\ are shown in Fig.~\ref{fig:momsCold12}. The kinematics of both phases of the cold gas are comparable. Exceptions are in the centre and perpendicularly to the jets in the NE, where blueshifted \HI\ is detected but no \CO\ is found. 

 The moment-2 maps (bottom panels of Fig.~\ref{fig:momsCold}) show that in most regions the velocity dispersion of both \HI\ and \CO\ gas are narrow ($\lesssim 50$\kms), as normally found in galaxy disks. In both phases of the gas, the highest dispersion is found where the radio jets bend in the NW and, more evidently in the SE. 
 
\section{The \Htwo/\HI\ ratio}
\label{sec:hihtwo}

\begin{figure}
	\centering
	\includegraphics[width=\columnwidth]{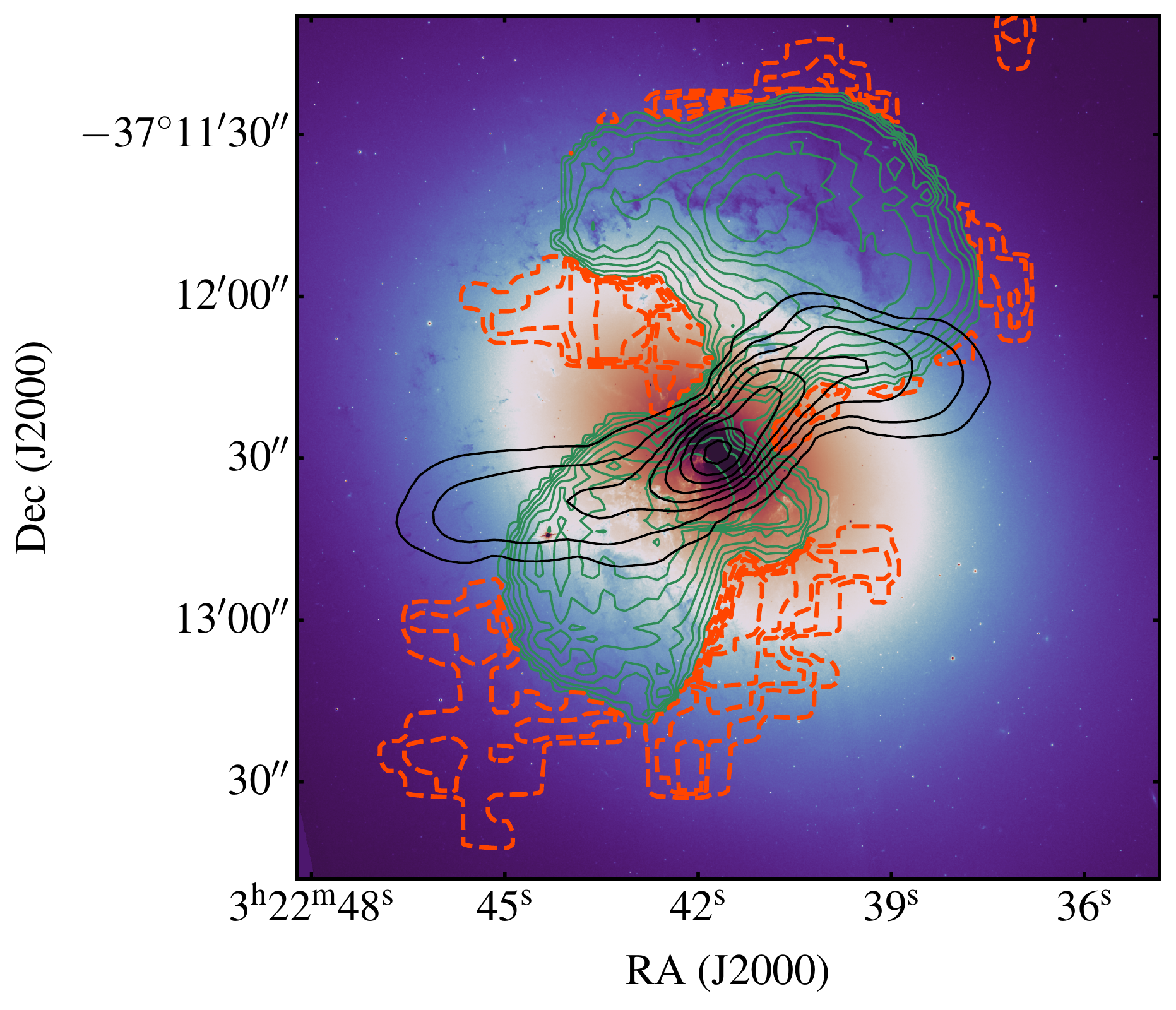}
	\caption{Hubble ACS visible emission (535 nm) of the centre of \forn. Green solid contours show regions where the inferred \Htwo\ mass surface density is greater than the \HI\ mass surface density (\Htwo/\HI$\gtrsim 1$), while orange dashed contours show regions rich in atomic hydrogen \HI, but where \Htwo\ is not detected. The radio jets are shown in black.}
	\label{fig:HI/Htwo}
\end{figure}

Fig.~\ref{fig:HI/Htwo} shows the \Htwo/\HI\ ratio distribution derived from the surface brightness maps. Green contours indicate a positive ratio, while orange contours show an over-abundance of \HI, and the ratio is determined from the $3\sigma$ detection limit of the molecular gas. The integrated \Htwo/\HI\ ratio is $\sim 5.5$, but this is given by averaging regions overabundant of molecular gas (as the outer arcs) with regions where only \HI\ is found. The \Htwo/\HI\ map reveals that in the outer arcs the \Htwo/\HI\ peaks (as high as $10$) correspond to the peaks of dust extinction. In the centre, along the nearly-edge on dust lane the \Htwo/\HI\ ratio is, on average, higher than in the outer arcs (6.8 and 3.5, respectively), while the regions perpendicular to the disk are depleted of \Htwo, and show ratios $\lesssim 1$. Along the jets (mainly in the south) we find a sharp increase of the \Htwo/\HI\ ratio while perpendicularly to the radio jets only \HI\ is found. In these regions, the \HI\ clouds have also very different projected velocity compared to the neighbouring regions (Fig.~\ref{fig:momsCold12}). These neutral gas clouds, with peculiar kinematics and no molecular counterpart, are also detected in the MUSE observations in the \nad\ absorption doublet~\citep[][]{Richtler:2020}. 

\begin{figure*}
	\centering
	\includegraphics[]{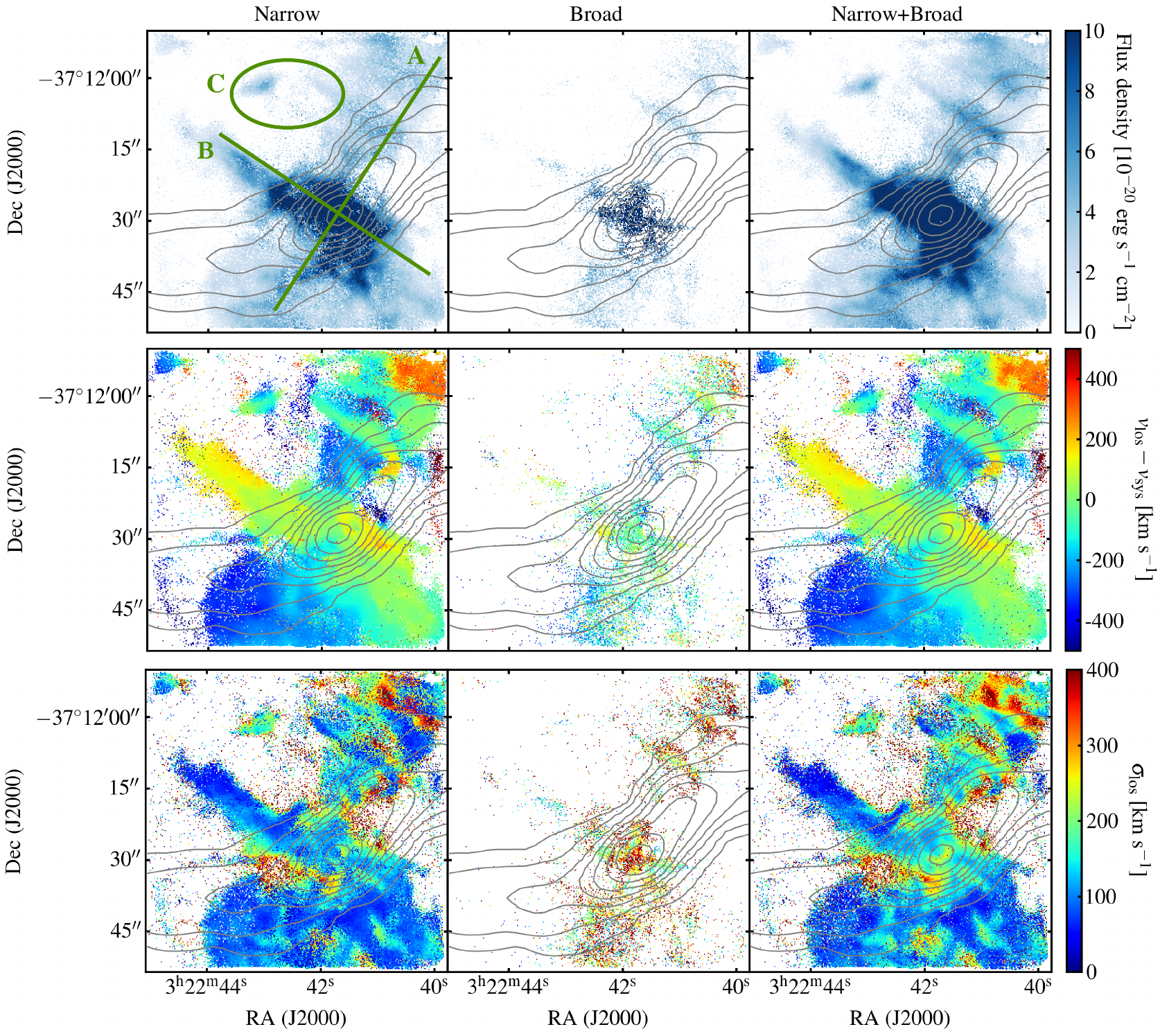}
	\caption{ Flux density distribution {(\em top row)}, velocity field {(\em central row)} and velocity dispersion maps {(\em bottom row)} of the ionised gas. The {\em left}, {\em middle} and {\em right columns} show the distribution of the narrow Gaussian component, of the broad Gaussian component and of the integrated fitted line (narrow+broad components), respectively. The radio jets of \forn\ are overlaid in grey in all panels. See Sect.~\	ref{sec:momsIon} for details on the regions identified in the top left panel.}
	\label{fig:momsIon}%
\end{figure*}

\section{Distribution and moment maps of the ionised gas in the innermost arc-minute}
\label{sec:momsIon}

From the results of the simultaneous fitting of all detected emission lines (see Sect.~\ref{sec:gaussFit}) we build the total flux distribution maps for each line (Fig.~\ref{fig:mom0Ion}).  A visual inspection of Fig.~\ref{fig:mom0Ion} suggests that the emission lines are mainly Low-Ionisation Nuclear Emission-line Region (LINER)-like, with notable exceptions in the centre and where the radio jets bend, where excitation appears to be more AGN-like.

This paper focuses on the kinematics of the multi-phase gas, hence and we will leave the discussion on the emission line ratios and excitation mechanisms to a following paper. 

The brightest line throughout the entire field of view is the \NIIsco. Since in the fitting procedure we constrained the centroid and width of all lines to be the same, we only show the properties of this line, which, in this study, is representative for all lines. In the top row of Fig.~\ref{fig:momsIon}, we show the line decomposition in Gaussian components. While most emission can be described using a single Gaussian (shown in the left column), the presence of a second broader component (central column) is required in the centre, and along filaments oriented approximately east-west, in particular, where the radio jets bend. In the top left panel of the Figure we highlight three peculiar distributions of the gas. The first (A-line) follows the overall distribution of the dust and cold gas along the NW-SE direction. The second (B-line) is oriented east-west (PA=80$^\circ$) and has no observed dust or molecular gas counterpart. We refer to this feature as the {\em EW-stripe}. Slightly to the north of this feature, we find several clouds of ionised gas with almost no dust (C-ellipse). These clouds coincide with the \HI\ clouds with highly blueshifted velocities and no molecular counterpart. In Appendix~\ref{app:specMUSE} we show different spectra extracted from the centre and these regions.

 In the central row of Fig.~\ref{fig:momsIon}, we show the velocity field of the ionised gas measured from the shift of the lines with respect to the systemic velocity of \forn\ (\vsys$= 1720$\kms). In the right column, where we show the sum of the narrow and broad components (left and middle column, respectively) the centroid is the velocity position of the barycentre of the line. Overall, the NW-SE distribution of the ionised gas appears to be dominated by rotation, as suggested by the redshifted velocities in the NW and blueshifted ones in the SE, and follows the same kinematics as the cold gas (Fig.~\ref{fig:momsCold12}). Notable exceptions are the EW filaments in proximity of the radio jets. These are redshifted ($v_{\rm los}\gtrsim 350$\kms) compared to the overall rotation of the neighbouring regions, and also their molecular counterpart (see Fig.~\ref{fig:momsCold}) was previously identified as non-rotating~\citep[][]{Morokuma:2019}.
 
The EW-stripe is aligned with the major axis of the stellar body and is perpendicular to the dust lane, which may lead to misinterpret the nature of this feature as an edge-on circumnuclear disk~\citep[\eg][]{Richtler:2020} as seen in several active early-type galaxies,~\citep[\eg\ NGC 1813 and PKSB~1718-649;][respectively]{Krajnovic:2015,Maccagni:2016,Maccagni:2018}. In \forn, however, the kinematics of the EW-stripe do not suggest that it can be an edge-on disk. The stripe has redshifted velocities in the east but does not show a blueshifted counterpart in the west. This is also visible in the single channel maps of the wide-field MUSE observations (see Sect.~\ref{sec:kinIon}).

In the last row of Fig.~\ref{fig:momsIon}, we show the velocity dispersion maps. To obtain the intrinsic dispersion of the line we deconvolve the measured dispersion ($\sigma_{\rm measured}$) by the spectral resolution of MUSE at the wavelength of the emission lines ($\Delta v_{\lambda}$)~\footnote{$\sigma_{\rm intrinsic}=\sqrt{\sigma_{\rm measured}^2-\bigg(\frac{\Delta v_{\lambda}}{2\sqrt{2\ln 2}}\bigg)^2}$}.

The broadest lines ($\sigma\gtrsim300$\kms) are detected in the centre and along the radio jets, where either a single broad and shallow component or two Gaussians best describe the observed lines. In Sect.~\ref{sec:feedback}, we further discuss the presence of outflowing gas in these regions.

\section{Analysis of the kinematics of the ISM in \forn}
\label{sec:kinIon}

We study the kinematics of the multi-phase ISM over the entire stellar body of \forn\ ($r \lesssim 10$ kpc) making use of the \HI\ cube, \couno\ cube  and of the \NIIsco\ datacube extracted from the MUSE mosaic (Sect.~\ref{sec:obsDR}). 

\begin{figure}
	\centering
	\includegraphics[trim = 0 0 0 0, width=0.99\columnwidth]{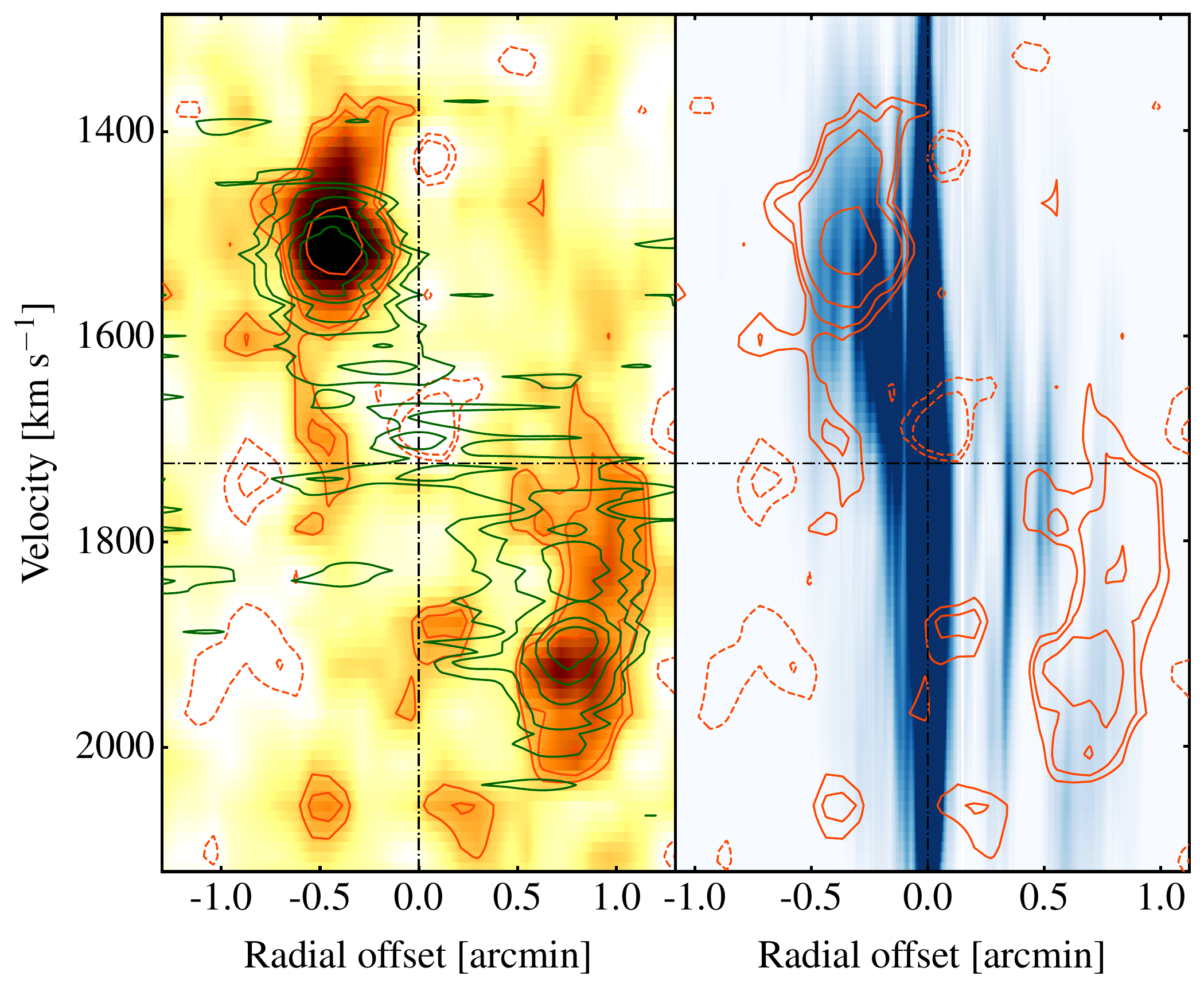}
	\caption{Position-velocity diagrams along the major axis of the inner nearly edge-on disk. {\em Left Panel:} Pv-diagram of the atomic and molecular gas (green contours). \HI\ emission is shown by the colour scale and orange solid contours while absorption is represented by dashed contours. Contour levels are $-5; -3; 3,5,6,12,24\sigma$. {\em Right Panel:} Pv-diagram derived from the \NIIsco\ emission (blue colour-scale) overlaid with the \HI\ emission and absorption.}
	\label{fig:pvPlotHICO}%
\end{figure}

The moment maps of the \HI\ and \CO\ suggest that in the innermost arcminute ($r \lesssim 6$ kpc) most gas rotates on a nearly edge-on disk. This is also supported by the appearance of the dust lane in the HST image (Fig.~\ref{fig:HST}), where most of the gas seen in extinction is located on the eastern side, suggesting that it is the nearest side. In the left panel of Fig.~\ref{fig:pvPlotHICO}, we show the position-velocity (pv-) diagram taken along the major axis of this disk (PA$=-30^\circ$). The smooth gradient of velocities of the \HI\ and of the \CO\ (left panel) confirms the overall rotation of the cold gas. Nevertheless, several clouds are deviating from the regular rotation both in the west and east side of the disk, as well as in the centre (previously identified in the moment maps, see Sect.~\ref{sec:momsIon}. In the centre, \HI\ is also detected in absorption in a single channel at the systemic velocity with optical depth of $\tau=0.004$ (which corresponds to $N_{\rm HI}\sim 3.0\times 10^{19}$\cmsq\ assuming a spin temperature of $100$ K). The right panel of Fig.~\ref{fig:pvPlotHICO} shows the pv-diagram of the \NIIsco\ line along the same axis, overlaid with the \HI\ gas. Most of the ionised gas also follows the overall regular rotation of the cold gas. Nevertheless, strong deviations are present throughout the entire disk and several clouds show counter-rotating velocities. In the centre the dispersion of the ionised gas is maximum, suggesting that the nuclear activity is still perturbing the gas in the innermost regions.

\begin{figure*}[t]
	\centering
	\includegraphics[trim = 0 0 0 0, width=\textwidth]{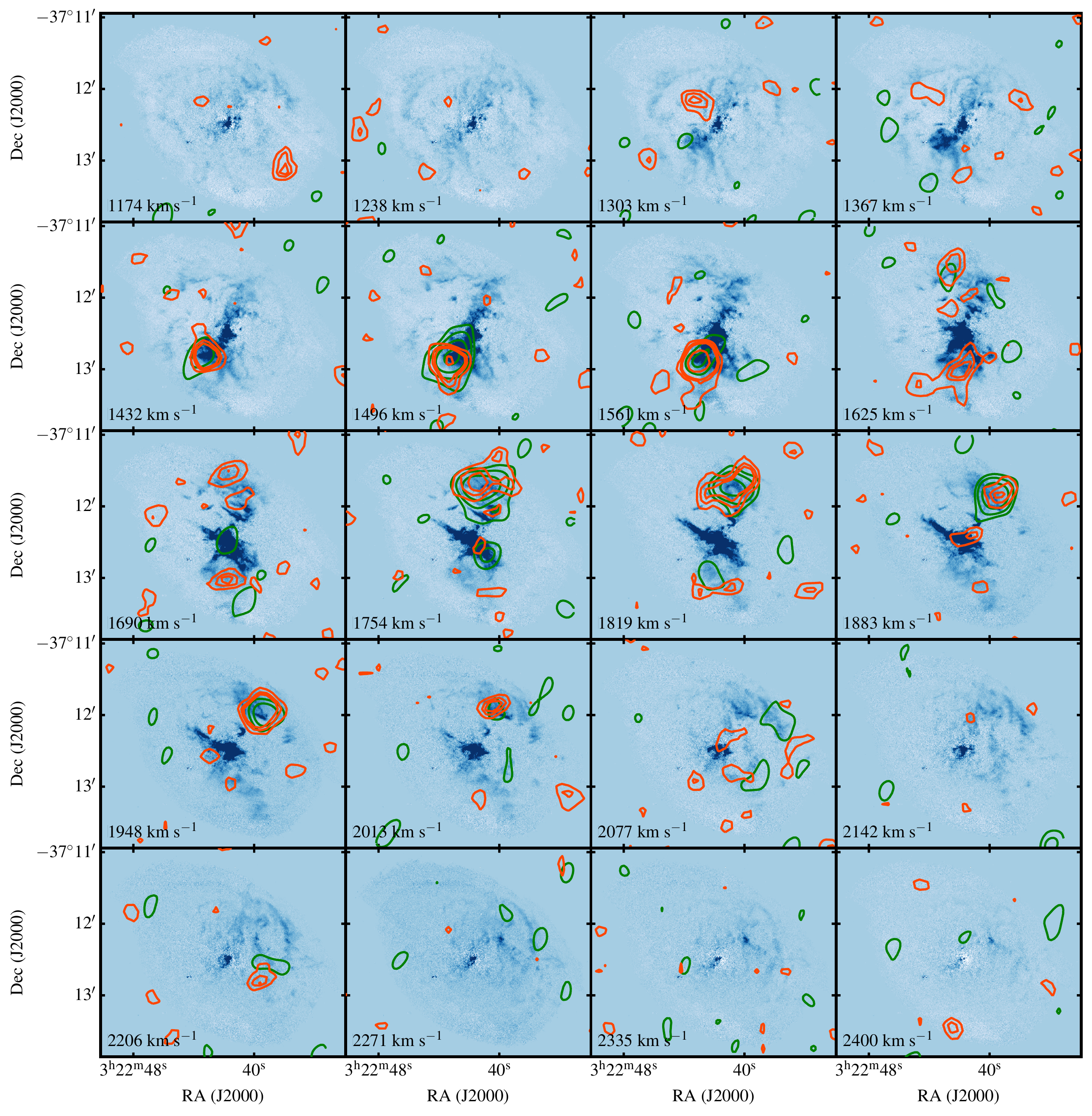}
	\caption{Consecutive channel maps of the \NIIsco\ line extracted from the MUSE wide-field datacube in the velocity range $1174$ -- $2400$\kms\ (the systemic velocity of \forn\ is $1720$\kms). The contours of the \CO\ emission ($0.03,0.06,0.12,0.15,0.2$ Jy beam$^{-1}$) and of the \HI\ gas ($0.15,0.3,0.6,1.2$~\mJyb) are shown in green and orange, respectively.}
	\label{fig:chanMapsAll}
\end{figure*}

Outside the innermost arcminute ($r \gtrsim 6$ kpc) the distribution of the ISM changes abruptly. The distribution of the dust (Fig.~\ref{fig:HST}) shows the presence of a  northern arc. The \HI\ and \CO\ gas follows the same arc and, therefore, cannot be co-planar with the inner, nearly edge-on disk. Furthermore, the velocity fields of the cold gas (top panels of Fig.~\ref{fig:momsCold12}) indicate that this arc is not just a more face-on extension of the inner disk because gas crosses the systemic velocity when moving from west to east along it (between approximately velocities 1600 and 2100\kms, see the emission marked by green and orange contours in Fig.~\ref{fig:chanMapsAll}) -- while, in the case of a simple inclination change, we would expect all gas in the northern arc to be redshifted, connecting in velocities with the northern edge of the inner disk.

The information provided by the cold gas is limited and we cannot know if the northern arc is part of a larger structure connecting the gas seen in the south-east of the inner disk. The ionised gas observations allow us to complete the picture. Fig.~\ref{fig:chanMapsAll} shows the single MUSE channel maps ($\Delta v\approx 75$~\kms) of the \NIIsco\ line in the velocity range $1174$--$2400$~\kms. The kinematics of the \NIIsco\ emission must be interpreted with caution. In the entire field of view, the line has large dispersion compared to the projected rotational velocity, and the blueshifted part of the line (channels $1174$ -- $1238$~\kms) overlaps with the \halpha\ emission. The ionised gas follows the kinematics of the \HI\ and \CO, showing that the northern (and southern) arc are part of an underlying, nearly circular distribution with radius $r\sim 6$ kpc (shown in grey in the $1238$\kms\ channel of Fig.~\ref{fig:chanMapsAll}), which from now on we call outer ring.

The channel maps highlight the following kinematical features of the ionised gas along the outer ring: gas at \vsys\ is located near the tips of the inner, nearly edge-on disk. Gas at the minimum and maximum velocity is located east and west of this inner disk. Between minimum (channels $v\lesssim 1400$\kms) and systemic velocity the ionised gas moves gradually along the outer ring from east towards the tips of the inner disk. Between systemic and maximum velocity (channels $v\gtrsim 2200$\kms) it moves gradually along the ring from the tips of the disk towards west. All this makes the kinematics of the outer ring resemble rotation. However, there are two problems with rotation. First, the apparent morphological and kinematical major axes of the outer ring are perpendicular to one another. The outer ring is slightly elliptical, with the long axis aligned approximately along the inner edge-on disk, and the short axis perpendicular to it. If there was rotation, we would expect to see the maximum and minimum velocities along the long axis (as for the inner disk, in the direction NE-SW). However, the terminal velocities are measured along the short axis. Second, even ignoring the first point above, the low ellipticity of the outer ring indicates its inclination is not nearly edge-on (as the inner disk) but more face-on ($i\lesssim 20^\circ$). Hence, the de-projected maximum velocity would imply a circular velocity of $\sim 2000$\kms. This is much larger than the expected circular velocity based on the baryonic Tully-Fisher relation: $v_{\rm rotational}=v_{\rm flat}=340$\kms~\citep[for $M_\star = 6\times10^{11}$\msun, ][]{Iodice:2017}. 

Both problems above are solved if we assume that the kinematics of the outer ring are due to radial motions. In this case, the morphological and kinematical major axes would be perpendicular to one another, as observed. For the outer ring to be a radial outflow (instead of a radial inflow) we need the east side to be the nearest to the observer, and the west side to be farthest. This is consistent with the fact that dust along the outer ring is more clearly visible on the east side in the HST image. \cite{Richtler:2020} also suggested the outflowing nature of this ring from the morphology and distribution of the dust and \NIIsco\ line, but detected the ionised gas in a smaller velocity range (between $\pm 300$\kms\ w.r.t. \vsys). Furthermore, expansion at $\sim 2000$\kms\ would be plausible. In this case it would have taken the outer ring $\sim 3$ Myr to expand from the centre to $\gtrsim 6$ kpc, which agrees with the total age of the radio jets~\citep[active plus non active phase][see Sect.\ref{sec:feedback} for further details]{Maccagni:2020}. 
 
\subsection{Modelling the kinematics of the ISM}
\label{sec:modISM}

To separate the inner rotating nearly edge-on disk and the outer radially expanding shell from the other components with non-rotating kinematics (see Sect.~\ref{sec:momsIon}) we build a parametric model. Given the complexity of the kinematics of the multi-phase gas and the clear deviations from regular rotation, we do not constrain all the parameters of the model with an automated fit. Instead, we build a schematic model~\citep[$^{3D}${\tt Bbarolo}\footnote{\url{https://editeodoro.github.io/Bbarolo/}}][]{diTeodoro:2015} that best describes the observed kinematics using different properties of the galaxy and the considerations made in the previous Sections.

 The velocity of the gas in the inner disk ($500$ pc $\lesssim r\lesssim 4$ kpc) is defined by the rotational velocity projected on the line of sight $v_{\rm rot}(r,i,{\rm PA})$ (which depends on the inclination and position angle of the disk), while the outer ring ($4\lesssim r\lesssim 8$ kpc) has only a radial velocity component $v_{\rm exp}(x,y,z)$. The inner disk has inclination ($i=89^\circ$) and position angle ($PA=-30^\circ$) constrained by the distribution of the dust, while the direction of rotation is constrained by the kinematics of the \HI\ and \CO. The disk has rotational velocity derived from the baryonic Tully-Fisher relation, $v^{\rm TF}_{\rm rot}\sim 340$\kms. In massive early type galaxies the rotation curve typically flattens in the innermost $500$ pc~\citep[][]{Noordermeer:2007,denHeijer:2015}, hence we consider a constant $v_{\rm rot}=340$~\kms\ throughout the disk. The expansion velocity $v_{\rm exp} = 2000$\kms\ of the outer ring follows from its low inclination and the large extent in velocities w.r.t. systemic ($\pm 700$\kms). We sample the two structures every $5$\arcsec. For simplicity, we constrain the velocity dispersion of each ring to be low, $35$\kms (approximately half the resolution of the MUSE observations). 

Figure~\ref{fig:chanMapsIonMod} shows the \NIIsco\ emission overlaid with the model in the velocity range $-2100,800$\kms\ with respect to \vsys. The model reproduces the overall kinematics of the ionised gas emission, except for the gas detected at the most redshifted and blueshifted velocities (channels $1174$~\kms\ and $2400$~\kms) in the centre and in the outer ring. The EW-stripe as well as the sub-filaments in the disk and gas in the centre stand out as deviating from the regular rotation. 
 
\subsection{The kinematics of the ionised gas in the innermost arcminute}
\label{sec:kIon}

As pointed out in the previous sections, in the innermost $\sim6$~kpc several gaseous clouds are deviating from the regular rotation of the nearly edge-on disk. In what follows, we use the model describing the gas kinematics over the entire stellar body to discriminate between the gas regularly rotating and the gas with irregular kinematics in the innermost arcminute. 

We define as `rotating', every spectrum of the \NIIsco\ datacube whose fitted emission line matches the model line for at least $50\%$ of its full-width at zero intensity. 

\begin{figure}
	\centering
	\includegraphics[trim = 0 0 0 0, width=\columnwidth]{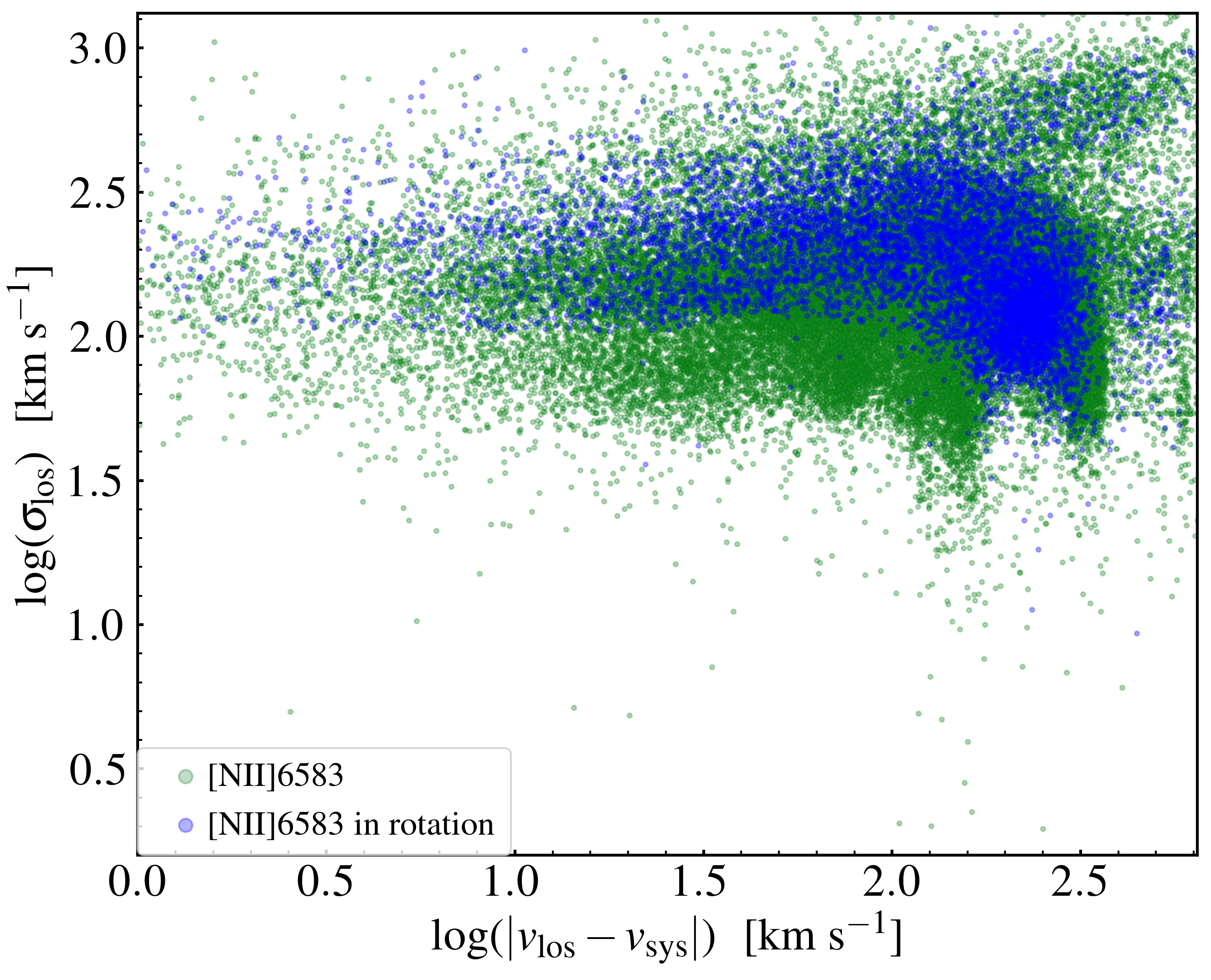}
	\caption{Velocity dispersion ($\sigma_{\rm los}$) vs. line shift with respect to the systemic velocity of the \NIIsco\ emission line ($|v_{\rm los}-v_{\rm sys}|$), in logarithmic scale. Pixels matching the modelled rotating disk are shown in blue, everything else is in green.}
	\label{fig:kplot}%
\end{figure}

\begin{figure*}
	\centering
	\includegraphics[keepaspectratio, width=\textwidth]{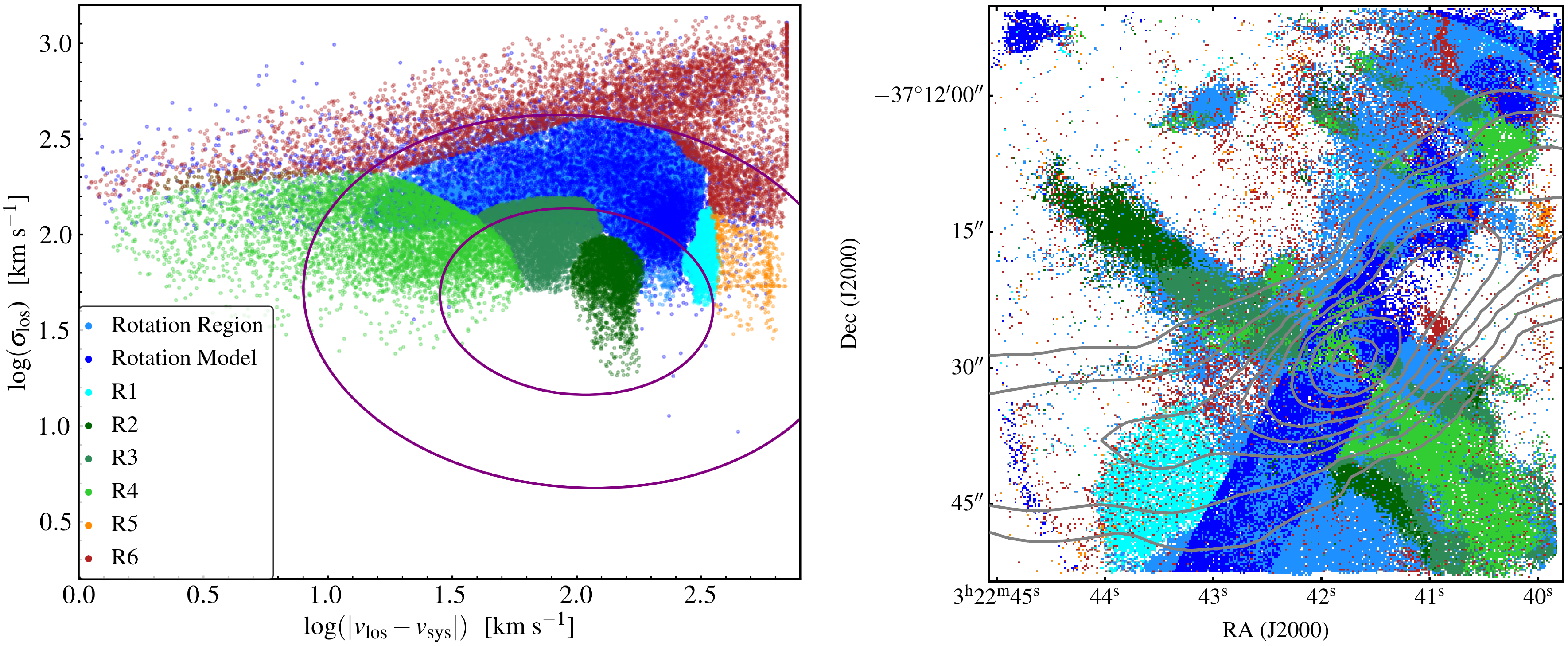}
	\caption{{\em Left Panel:} velocity dispersion  vs. line shift (k-plot) with respect to the systemic velocity of the \NIIsco\ line emission in logarithmic scale. Colours identify different regions in the plot. Pixels consistent with the model of rotation are shown in blue, while pixels in the kinematical region of rotation are shown in light blue. The purple ellipses shows the $1,2\sigma$ confidence intervals tied to the global log-normal distributions found for chaotic cold accretions~\citep[][]{Gaspari:2018}. {\em Right Panel:} Distribution of the ionised gas. Colours correspond to the plot in the left panel. Radio jets are shown in black contours.}
	\label{fig:Kreg}%
\end{figure*}

Following \citet{Gaspari:2018}, Fig.~\ref{fig:kplot} (the kinematical-plot, `k-plot') shows the line width of the total \NIIsco\ line\footnote{since the total line is made of multiple components, we compute the line-width as the dispersion of the total fitted line} vs. its velocity centroid (with respect to the systemic velocity of the galaxy) for each independent line of sight detected by MUSE in the innermost arcminute of \forn\ (green colours). Blue dots are the spectra matching the regular rotation. Remarkably, the rotating spectra mark a well defined region in the k-plot, with average width $\sim200$\kms and the centroids spanning all values, with most points accumulating at $\pm 220$\kms. The left panel of Fig.~\ref{fig:Kreg} shows that also the non-rotating points have clustering patterns: {by looking at their density in the k-plot, six different loci can be identified.} Their spatial distribution is shown by the same colours in the right panel of the Figure. The gas with large line widths ($R6$, red colours) well matches the regions previously identified with highest velocity dispersion (bottom row of Fig.~\ref{fig:momsIon}). The spectra with similar widths to the rotating gas but extreme centroids are all in the SE wake of the radio jets ($R1$, cyan colours), or in the NW bend of the jet ($R5$, orange colours), suggesting that also this gas may be outflowing (see Sect.~\ref{sec:feedback} for further details). 

 The gas with systematically narrower width than the rotating gas appears mostly distributed along the EW-stripe ($R2$, $R3$ and $R4$, green colours), and in the sub-filaments of the edge-on disk. In Sect.~\ref{sec:feeding} we discuss the nature of these filaments according to their kinematical properties.
 
 In Table~\ref{tab:masses} we show the masses of the ionised gas for each region. The mass of the ionised gas is computed as in, for example,~\cite{Poggianti:2019}: 

\begin{equation}
M_{\rm ion} = \frac{L_{H\alpha}\,\,m_H}{n\,\,\alpha_{H_\alpha}\,\, h\nu_{H\alpha}} 
\end{equation}
 
 where $L_{H\alpha}$ is the luminosity of the \halpha\ line, corrected for dust extinction estimated from the Balmer decrement~\citep[\eg][]{Dominguez:2013}. $n$ is the gas density computed from $R=$\SIIdue/\SIIuno, following the calibration of~\citealt{Proxauf:2014} (which is valid for $0.4\leq R\leq 1.4$). $m_{\rm H}$ is the mass of the hydrogen atom, $\alpha_{H\alpha}$ is the effective \halpha\ recombination coefficient, and $h\nu_{H\alpha}$ is the energy of the \halpha\ photon.
  
\subsection{The kinematics of the cold gas}
\label{sec:kCold}

\begin{figure*}
	\centering
	\includegraphics[keepaspectratio, width=\textwidth]{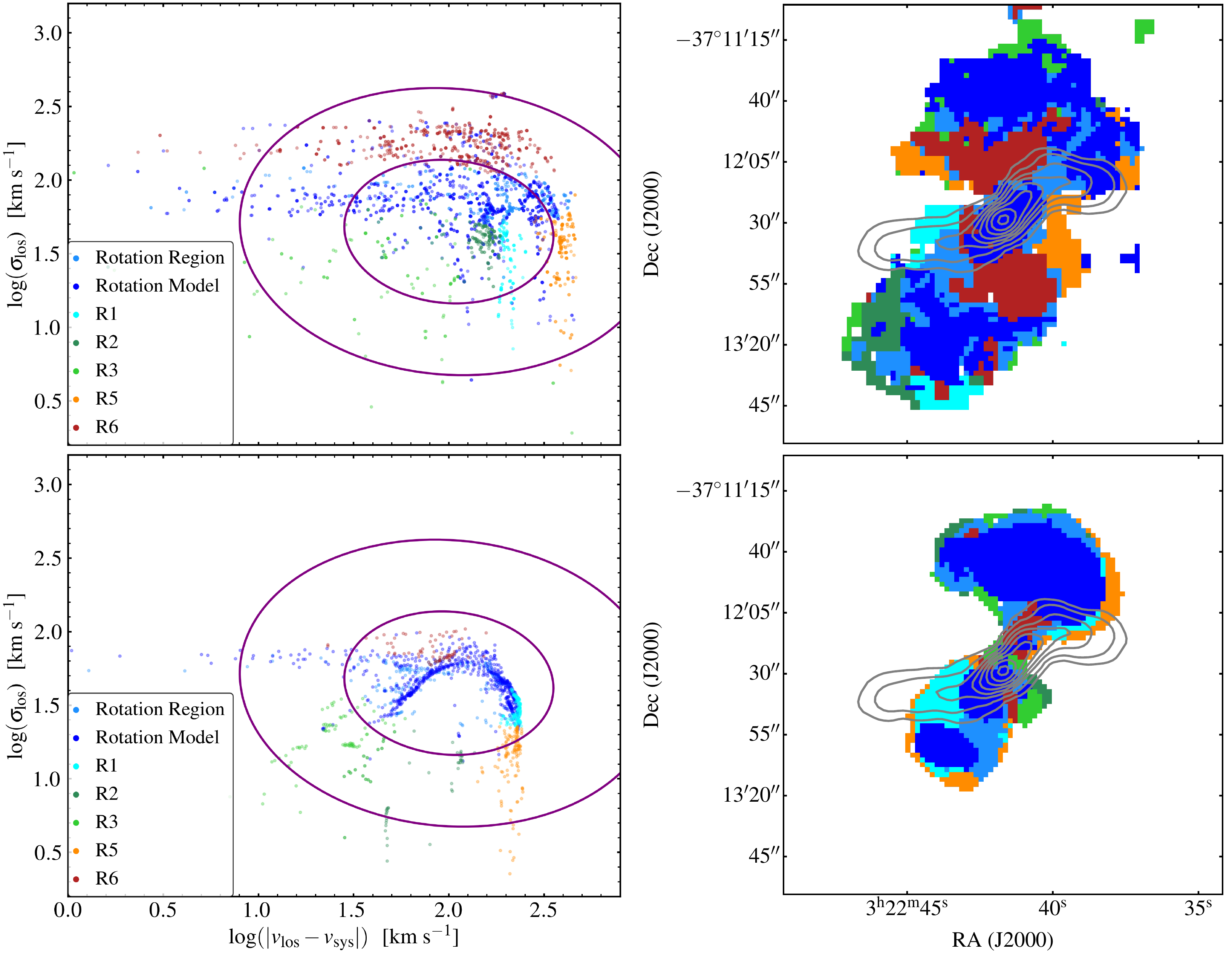}
	\caption{{\em Left Panels:} velocity dispersion vs. line shift with respect to the systemic velocity of the \HI\ and \CO, respectively, in logarithmic scale. Colours identify different regions in the plot. Pixels consistent with the model of rotation are shown in blue, while pixels in the kinematical region of rotation are shown in light blue. The purple ellipses are as in the left panel of Fig.~\ref{fig:Kreg}. {\em Right Panels:} Distribution of the \HI\ and \CO, respectively. Colours correspond to the plot in the left panel. Radio jets are shown in grey. The \HI\ and \CO\ different loci, in particular $R6$ and the rotating regions are distributed similarly to the ionised gas regions, on a larger scale.}
	\label{fig:KregHI}%
\end{figure*}

We apply the same procedure described in the previous Section to study the kinematics of the cold gas. We identify components matching with the regular rotation similarly to the ionised gas. Because of the wide-field of view of the cold gas observations, we include in this analysis also the outer ring. In Fig.~\ref{fig:KregHI} we identify in the k-plots of the \HI\ (top panel) and \CO\ (bottom panel) the same loci that we identified for the ionised gas. The right panels of the Figures show their spatial distribution. Compared to the ionised gas, we have fewer independent points because of the coarser spatial resolution of the observations. Most loci correspond to the ones identified in the ionised gas\footnote{except for $R4$ that having too few points is included in $R3$}. The large width and centroid ($R1$, $R5$, $R6$; cyan, orange and red colours, respectively) are located in the wake of the radio jets and in the outer rings. The narrow line-widths ($R2$, $R3$ green colours) of the \CO\ are found perpendicular to the edge-on disk. The \HI\ and \CO\ masses of each region are shown in Table~\ref{tab:masses}.  Even though the line-widths are intrinsically smaller, similarly to the ionised gas the rotating \HI\ and \CO\ identify a similar region in the k-plot (with most points around $|v_{\rm los} - v_{\rm sys}|\sim 224$~\kms, $\sigma_{\rm los}\sim 80$~\kms). The non-rotating points are also clustering in the region with high sigma and centroid ($\sigma_{\rm los}\gtrsim 125$~\kms, $|v_{\rm los} - v_{\rm sys}|\sim 224$~\kms).  

\section{Discussion}
\label{sec:discussion}

We relate the kinematical properties of the loci identified in the k-plot of the ionised and cold gas to their impact on the AGN activity.

\begin{table}
        \caption{Masses of the different gaseous features identified in \mbox{NGC~1316}}
        \centering
        \label{tab:masses}
        \begin{tabularx}{\columnwidth}{X c c c c c}  
                \hline\hline                                                         
                Region	& $M_{\rm HI}$ [\msun]  & 	$M_{\rm H2}$ [\msun]	& $M_{\rm ionised}$ [\msun] \\ 
       	        \hline
                Rotating	&	$4.9\times 10^7$    & 		$4.7\times10^8$		&  $1.1\times10^6$  \\
                $R1$		&	$9.1\times 10^5$    & 		$5.1\times10^7$		&  $4.1\times10^4$ 	\\  
                $R2$		&	$1.8\times 10^6$    & 		$2.6\times10^6$		&  $3.6\times10^4$  \\      
                $R3$		&	$1.5\times 10^6$    & 		$7.0\times10^6$		&  $8.8\times10^4$	\\     
                $R4$		&	-      				& 		-					&  $2.3\times10^5$	\\
                $R5$		&	$2.5\times10^6$     & 		$7.1\times10^6$		&  $1.1\times10^3$ 	\\
                $R6$		&	$1.9\times10^7$     & 		$2.4\times10^5$		&  $1.1\times10^5$ 	\\
                \hline
                Total  	&	$6.7\times10^7 $    & 		$5.8\times10^8$		&  $1.6\times10^6$	\\					 
                \hline                           
        \end{tabularx}
        \tablefoot{For further information about the definition and distribution of the regions refer to Sec.~\ref{sec:kinIon}.}
\end{table}

\subsection{Feeding: raining multi-phase clouds and filaments via turbulent condensation}
\label{sec:feeding}

The k-plots shown in the previous sections indicate that the kinematical properties of the multi-phase gas in \forn\ can be broadly classified in three groups, \ie:
\begin{itemize}
	\item gas rotating within the inner disk (blue colours, in Figs.~\ref{fig:Kreg},~\ref{fig:KregHI})
	\item gas distributed  in the EW-stripe and filaments perpendicular to the rotating disk with narrow line-widths ($R2$, $R3$ and $R4$; green colours)
	\item gas distributed in clouds in the wake of the radio jets and in the outer ring with large line-widths ($R1$ in cyan and $R5$, $R6$ in red colours)
\end{itemize}

The physical properties of the cold gas in these regions are also very different (see Sect.~\ref{sec:hihtwo}), \ie\ the rotating regions have highest \Htwo/\HI\ ratio, while this ratio decreases in the outflow and is $\lesssim 1$ in the EW-stripe.

The EW-stripe and filaments have the most peculiar morphology and kinematical properties. The EW-stripe has kinematics in conflict with a rotating circumnuclear disk, as indicated by the presence of redshifted velocities w.r.t. \vsys\ on both sides of the centre. These shifts are small and, along with the small velocity dispersion of the stripe, suggest it is not outflowing. The sub-filaments are spatially connected with the edge-on disk but their kinematics are very different. They are highly redshifted compared to the overall rotation of the disk, and have narrower line widths. 

The k-plot gives us a hint on the nature of the EW-stripe and filaments. The purple ellipses in the left panels of Figs.~\ref{fig:Kreg},~\ref{fig:KregHI} show the mean value of line widths and shifts $\pm 1, 2\sigma$ expected from CCA simulations, if the cold and ionised gas were undergoing condensation and inducing rainfalls~\citep[see Fig.~4 of][]{Gaspari:2018}. 
In the ionised and \HI\ gas the EW-stripe and filaments are consistent with the expectations of CCA. The k-plot also shows that the outer edges of the EW-stripe ($R2$) are more compatible with CCA than the inner regions ($R3, R4$). Possibly because, the core of \forn\ may have been recently re-activated~\citep[][]{Maccagni:2020} and that the innermost regions are perturbed by its activity (as it is suggested by the presence of a broad component, $\sigma\gtrsim 300$\kms\ in the ionised gas emission lines). 
 
 \begin{figure}
	\centering
	\includegraphics[trim = 0 0 0 0, width=\columnwidth]{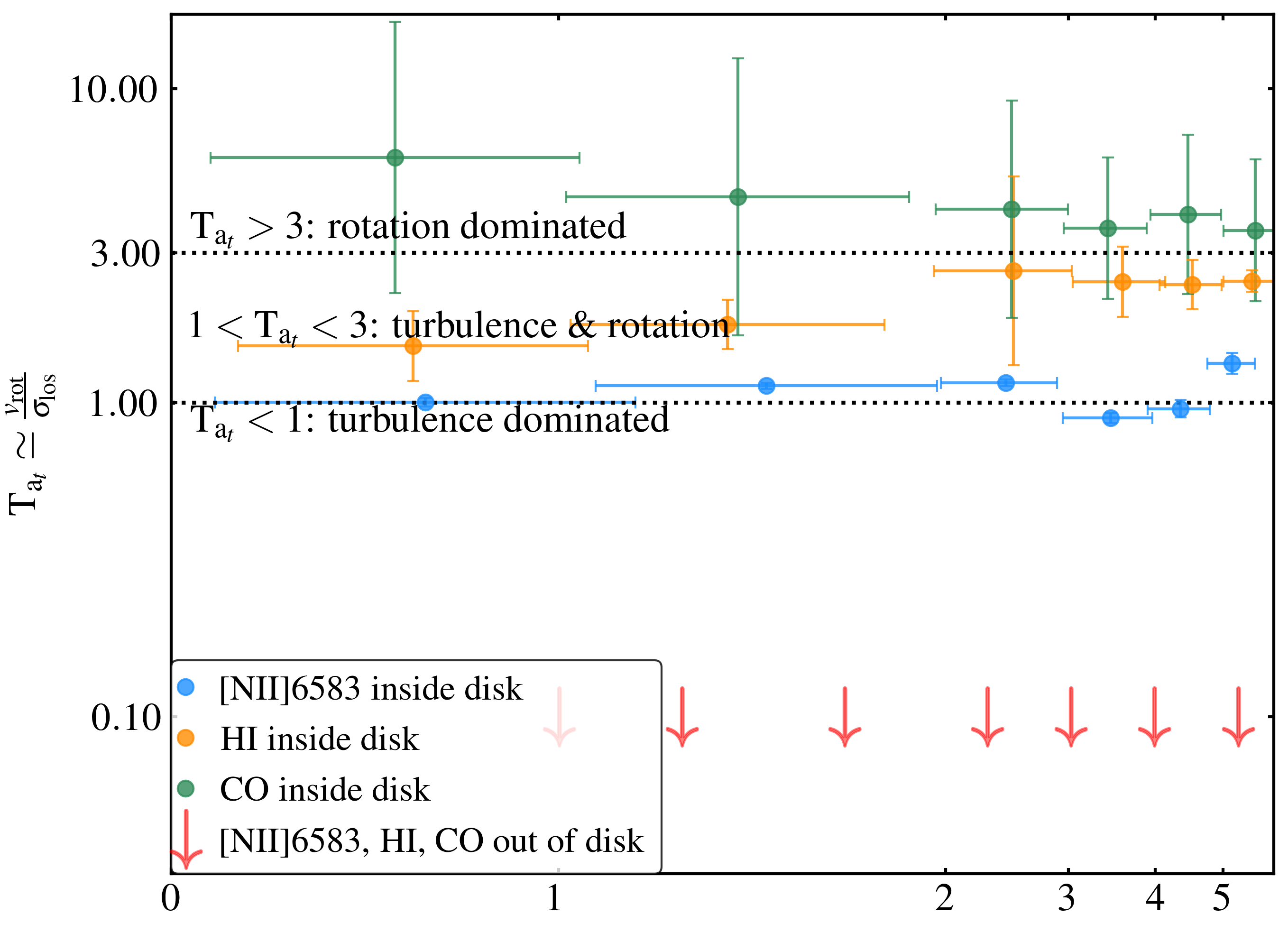}
	\caption{Radial profile of the Taylor number (${\rm Ta_t} = {v_{\rm rot}}/{\sigma_{\rm los}}$; \citealt{Gaspari:2015_cca,Gaspari:2017_cca}), in logarithmic scale. In green, orange and blue colours we show three phases of the gas within the rotating disk, in red the upper limits to ${\rm Ta_t}$ of the EW-stripe and filaments outside the disk, which are mostly dominated by pure turbulence. 
	We label the areas of rough transition from rotation- to turbulence-dominated (`CCA-driven') kinematics, with dotted lines as separators.}
	\label{fig:taylorNumber}
\end{figure}
 
 The turbulent Taylor number Ta$_{\rm t} \equiv v_{\rm rot}/\sigma_{v, \rm 3D}$ (\citealt{Gaspari:2017_cca})\footnote{$\sigma_{v,\rm 3D}=\sqrt{3}\sigma_{\rm los}$ is the three-dimensional velocity dispersion.} is a key dimensionless number  that compares the rotating and turbulent kinematics. Typically, when Ta$_{\rm t}< 3$ turbulence becomes significant compared to rotation and chaotic condensation may occur because of it. In particular below the unity threshold, CCA becomes vigorous inducing extended filamentary condensation, instead of a more disk-dominated condensation (\citealt{Gaspari:2015_cca}).

Here, we use the Taylor number to determine the role of turbulence of the multi-phase ISM in the centre of \forn. In the EW-stripe and filaments, since their kinematics is not compatible with rotation, $v_{\rm rot}\to 0$ and Ta$_{\rm t}\to 0$. Even though the gas in the inner disk is dominated by rotation, its velocity dispersion ($\sigma_{\rm los}$) is high in all phases of the gas (as discussed in the previous Sections). Hence, we compute Ta$_{\rm t}$ for all phases throughout the disk. From the rotating model (Sect.~\ref{sec:modISM}), we fix  $v_{\rm rot}=340\sin(i)$\kms, while we measure $\sigma_{\rm los}(r)$ as the mean value in radial bins of $1$ kpc. The results are shown in Fig.~\ref{fig:taylorNumber}. We highlight with dotted lines and labels where turbulence becomes significant, starting to perturb the rotating disk.
 
 The Figure shows that the ionised and (inner) \HI\ gas phases inside the disk are still altered by turbulence. Several clouds in these phases may be interacting with in-falling clouds, which perturbs the coherent rotational structure, hence driving a clumpy disk. The ionised gas has always ${\rm Ta_t} \sim  1$, \ie\ the turbulent component mildly dominates over the rotational velocity. Thus, we expect strong clumpyness and/or filaments of this gas (and related boosted CCA feeding), as unveiled by the emission maps (Figs.~\ref{fig:momsIon} and~\ref{fig:mom0Ion}) and k-plot (Fig.~\ref{fig:Kreg}) in a consistent manner. The molecular component in the disk appears to reside in the pure rotational regime, as it is physically the densest and thus most compact phase, which is difficult to internally perturb via turbulent motions~\citep[\eg][]{Gaspari:2017_cca}.

Further constraints on the feeding nature of the EW-stripe and filaments can be given by the radial variation of the $C$-ratio in the different phases of the gas. The $C$-ratio ($C\equiv t_{\rm cool}/t_{\rm eddy}$) is the criterion closely related to CCA and ensuing top-down condensation cascade~\citep[][]{Gaspari:2018}. When the eddy turnover time-scale ($t_{\rm eddy}=2\pi r^{1/3}L^{2/3}/\sigma_{v, \rm 3D}(r)$) is comparable to the cooling time, extended filamentary and cloudy condensation can efficiently occur. 
The cooling time is defined from the properties of the X-ray halo in which the galaxy is embedded, $t_{\rm cool}=3k_b T/n_e\Lambda$~\citep[where $\Lambda(T,Z)$ is the plasma cooling function;][]{Sutherland:1993}. The $C$-ratio has already been used to link the level of thermal instability of the ISM, Intra-Group Medium and ICM to the on-going nuclear activity of different AGNs, such as Abell 2597~\citep[][]{Tremblay:2018}, and NGC 7409~\citep[][]{Juranova:2019}, and a dozen other clusters~\citep[][]{Olivares:2019}. 

We estimate the cooling time in the centre of \forn\ from the temperature of the X-ray halo measured by~\cite{Nagino:2009} between 1 and 8 kpc, $\langle T_{\rm x} \rangle = 0.77$~keV. Since the turbulence velocities of the ionised gas linearly correlate (with a scatter of $14\%$) with the velocities of the X-ray emitting gas~\citep[][]{Gaspari:2018}, we determine the $C$-ratio of three different regions of \forn\ from the kinematics of the ionised gas. The regions are selected from the k-plot: the gas that is rotating, the gas likely outflowing ($R1$, $R5$ and $R6$) and the gas with kinematics compatible with the expectations of CCA (the EW-stripe and filaments $R2$, $R3$, $R4$). To determine the $C$-ratio, we measure the mean values of the velocity dispersion in radial bins of $1$ kpc along the selected regions. 

\begin{figure}
	\centering
	\includegraphics[keepaspectratio, width=\columnwidth]{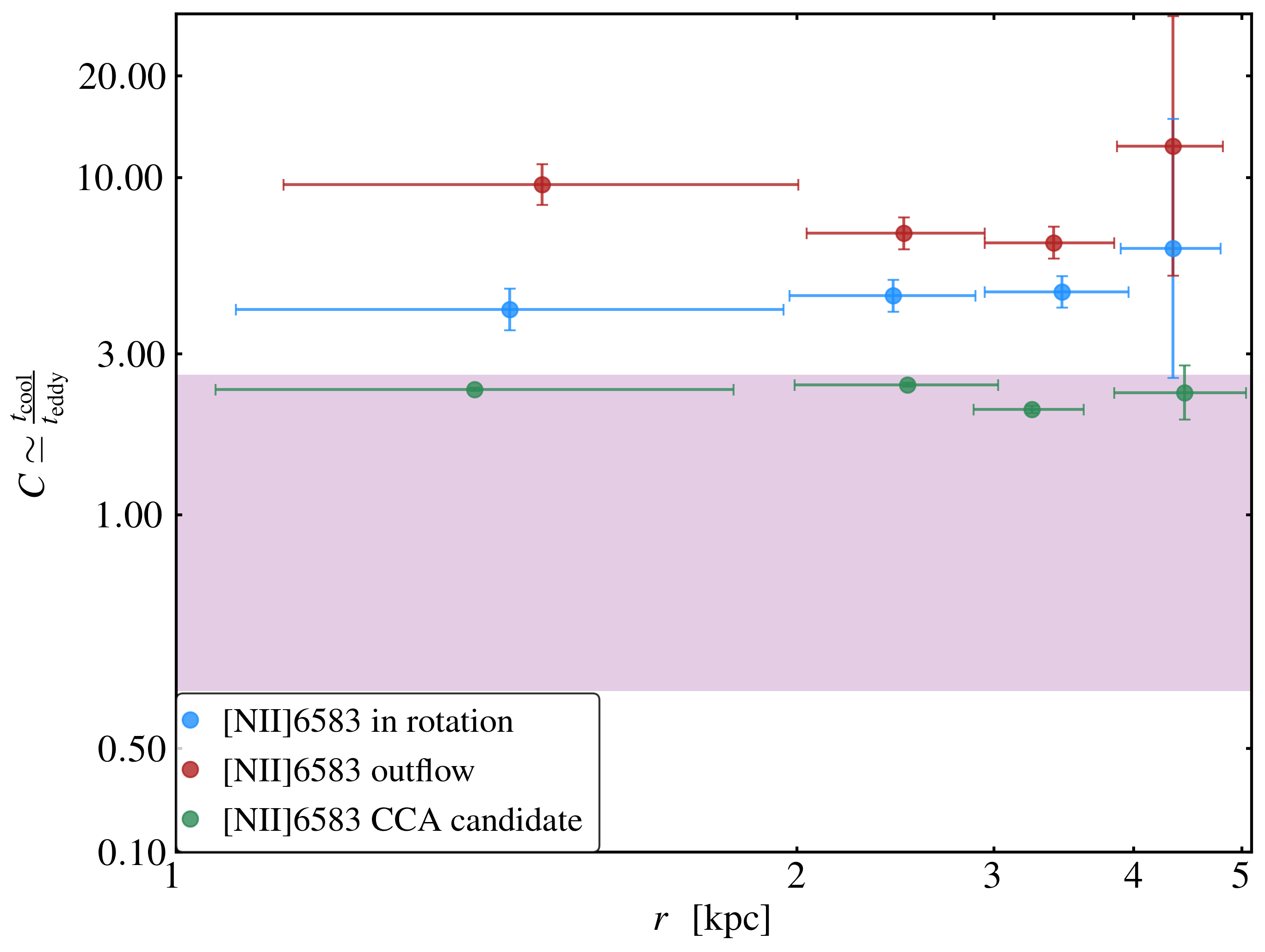}
	\caption{Radial profile of the $C$-ratio ($t_{\rm cool}/t_{\rm eddy}$) in logarithmic scale. The purple-shaded area shows where the gas becomes thermally unstable, leading to extended filamentary condensation and the CCA rain (e.g.~\citealt{Gaspari:2018}).}
	\label{fig:cRatio}
\end{figure}

The results are shown in Fig.~\ref{fig:cRatio}. The purple band indicates when the gas is being driven by nonlinear thermal instabilities (in particular around unity value). The outflowing and rotating regions have $C$-ratio significantly above unity, hence the multi-phase condensation is weak, if not delayed. Conversely, the CCA regions are those with the lowest $C$-ratio approaching unity, especially at intermediate and large radial distances. The most thermally unstable region is the EW-stripe, which is consistent with the above inspection of the maps and k-plot. This stripe is also perpendicular to the outflow/disk region, as expected in typical self-regulated CCA/feeding and bipolar outflow/feedback cycles \citep[e.g.~][]{Gaspari:2012b}.  The mass of the EW-stripe is $3.5\times10^5$~\msun\ for the ionised gas, $4.3\times10^6$~\msun\ for the \HI\ and $9.6\times10^6$~\msun\ for the \Htwo\ derived from the \CO\ observations. The latest episodes of nuclear activity of \forn\ were rapid~\citep[1-3 Myr;][]{Maccagni:2020} and emitted energies typical of low excitation radio galaxies ($\lambda = (L_{\rm mech}+L_{\rm rad})/L_{\rm Edd}\sim10^{-3}$). Under low-efficiency accretion regimes ($\dot{M}\lesssim 0.1$~\msunyr) the mass of the filaments is sufficient to fuel the nuclear activity.

The above k-plot, together with the Taylor number and the $C$-ratio constraints, as well as the rapid flickering in the low-power regime, show that CCA is very likely on-going within the EW-stripe. There, gas is likely condensing from a diffuse ionised phase to neutral and then to molecular clouds, thus fuelling the nuclear activity of \forn. 

 Besides its kinematical properties, the EW-stripe also has physical differences compared to the gas in the centre and the outer ring. As shown in Fig.~\ref{fig:mom0Ion}, the stripe is clearly detected in the low-ionisation lines (\SIIdoublet, \NII, \halpha) but is much weaker in the high-ionisation lines (\hbeta, \OIII, and \OI\footnote{not shown in the Figure}). Given that the high ionisation lines are better tracers of shocks and heating processes, this further suggests the condensing nature of the EW-stripe compared to the other regions identified in the k-plot, since feeding proceeds more abundantly when heating is low and cooling high (this is also why feeding seems more pronounced perpendicular to the jet/feedback region). The EW-stripe is rich of ionised gas and \HI\ but lacks of molecular gas (Sect.~\ref{sec:hihtwo}) and dust (Fig.~\ref{fig:HST}). Plausibly, even though accreting, the stripe has not fully reached the final phase of condensation yet. 

\subsection{Feedback: outflowing gas in the wake of the radio jets}
\label{sec:feedback}

\begin{figure}
	\centering
	\includegraphics[trim = 0 0 0 0, width=\columnwidth]{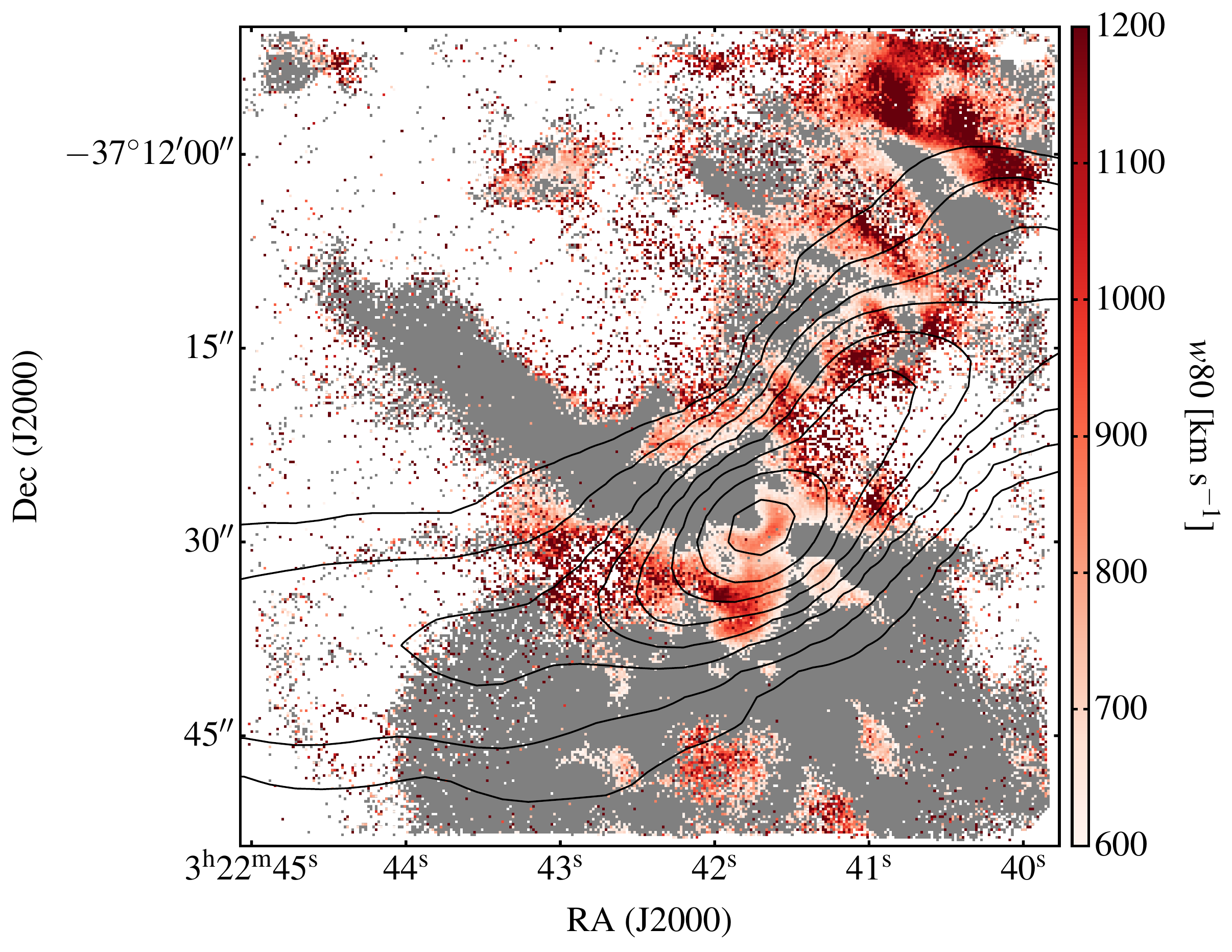}
	\caption{Map of the $w80$ of the total \NIIsco\ line. Values above $600$\kms\ may indicate the presence of outflowing gas, and are shown in colour-scale. The radio jets are shown in black contours.}
	\label{fig:w80}%
\end{figure}

The velocity fields and dispersion maps of the cold and ionised gas (Figs.~\ref{fig:momsCold12} and ~\ref{fig:momsIon}, respectively), as well as the k-plot of the multi-phase ISM (Figs.~\ref{fig:Kreg},~\ref{fig:KregHI}), show that in the centre and along the radio jets of \forn\ in different regions ($R1$, $R5$ and $R6$ in the k-plots) the gas has extreme line-widths and/or is highly shifted w.r.t. \vsys, suggesting the presence of an outflowing component. For a direct comparison with previous works on outflowing gas in AGN ~\citep[see, for example,][]{Harrison:2016,Wylezalek:2020,Kakkad:2020} we adopt as measurement of these extreme line-widths, the width at $80\%$ of the total flux of the \NIIsco\ line, $w80$. Its distribution is shown in Fig.~\ref{fig:w80}. The colour-scale highlights the regions where the $w80\gtrsim600$~\kms, which is a typical velocity cutoff used to define gaseous outflows. The figure shows the filamentary nature of the outflows, and the close correspondence with the main features of the radio jets: the core, the wake in the innermost kilo-parsec and the bends at $\sim4.5$ kpc from the centre. 

\begin{figure*}
	\centering
	\includegraphics[keepaspectratio,width=0.9\textwidth]{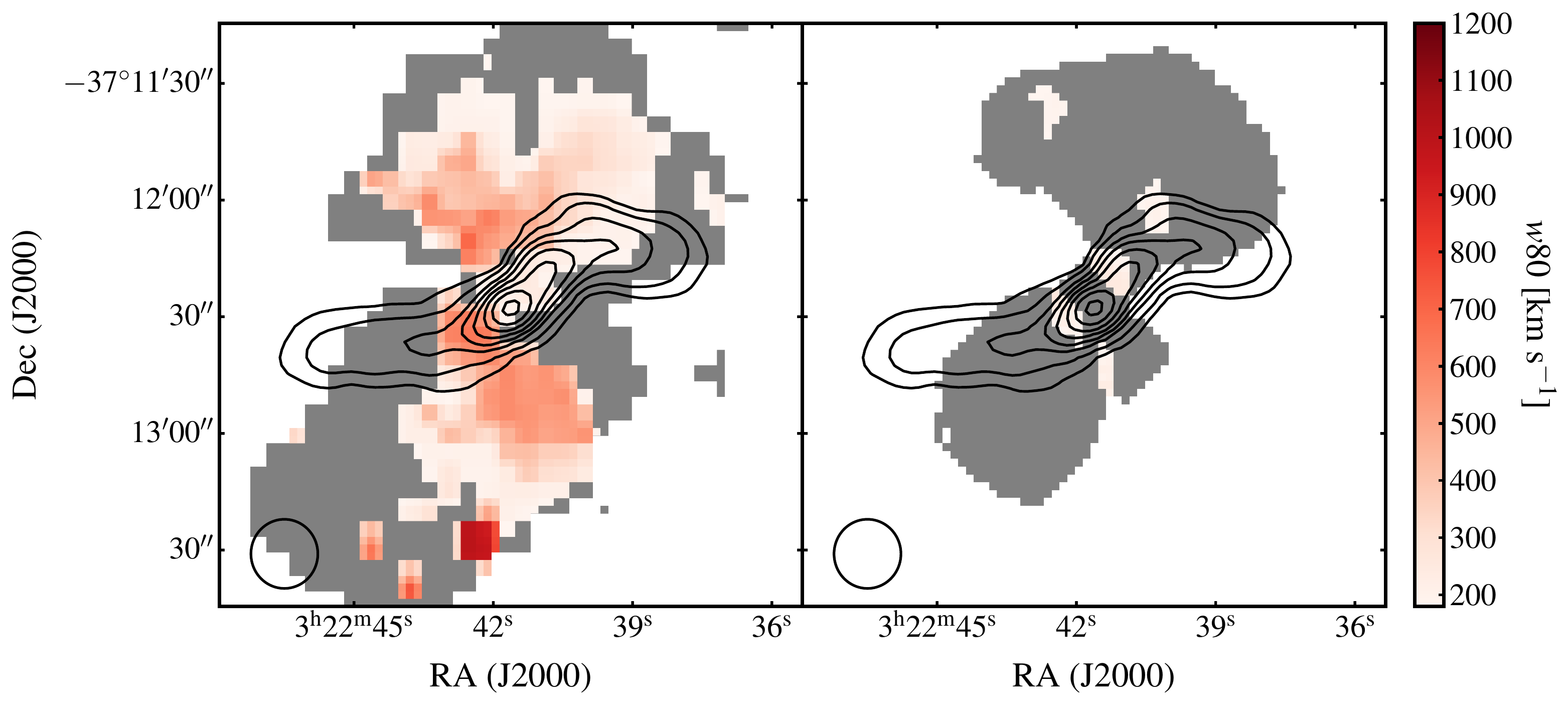}
	\caption{{\em Left Panel}: Map of the $w80$ of the \HI\ line. Regions with broad emission $w80\gtrsim200$\kms\ are coincident with regions where ionised gas may be outflowing. {\em Right Panel}:  Map of the $w80$ of the \couno\ line. Regions with broad emission $w80\gtrsim200$\kms\ are coincident with regions where ionised gas may be outflowing.}
	\label{fig:w80coldGas}%
\end{figure*}

Figure~\ref{fig:w80coldGas} shows that where the ionised gas seems outflowing also the cold component has extreme kinematics ($w80\gtrsim200$\kms)\footnote{In the \meer\ and ALMA observations, the shallow broad wings that dominate $w80$ are often seen below the $S/N\sim3$, making uncertain a direct measurement of the full extent of the lines. Hence, we infer $w80$ from the dispersion of the line assuming the relation for a single Gaussian profile ($w80=2\sqrt{2\ln 2}\sigma /0.919$).}. Because of the wider field of view of the \HI\ and \CO\ observations, we notice that large line-widths are present in the ring, too, further suggesting the presence of a multi-phase outflow out to $6$ kpc from the AGN. Given that to reach this distance the gas in the outer ring may have taken the same time as the age of the radio jets ($3$ Myr at $v_{\rm OUT}\sim 2000$\kms, see Sect.~\ref{sec:kinIon}), the turbulent kinematics of the outer ring were possibly caused by the same nuclear activity that generated the jets. 

Outflowing gas extending for several kilo-parsecs in AGNs are typically detected in a bi-conical distribution, perpendicularly or along the jets\citep[as in, for example, NGC 1266 and NGC 1068, see][respectively]{Alatalo:2011,Garcia-Burillo:2019}, and present radial velocities of several thousands of \kms. In \forn\ instead we detect a ring-like distribution of the gas with radial velocities not as extreme ($\sim 2000$\kms). A possible explanation may lie in the history of the nuclear activity of \forn. Since, currently, the jets are not active, but have been adiabatically expanding through the ISM for the last 2 Myr, possibly, we are seeing a post-outflowing phase where the kinematics are still turbulent, but the AGN winds are not accelerating nor exciting the gas anymore. This would also explain the LINER-like excitation rates of the outflowing material.

Since the $w80$ maps are compatible with the results of the k-plot, we use the masses of the \HI, \CO\ and ionised gas in regions $R1$, $R5$ and $R6$ (see Table~\ref{tab:masses}) to determine the total mass of the outflow~$M_{\rm out}\sim 2\times10^7$~\msun. Assuming that the outflow was ejected at the same time of the jets, the mass outflow rate is $\dot{M}_{\rm OUT}\sim 5$~\msunyr. 

The (macro) outflowing power is $P_{\rm OUT}=(1/2)\dot{M}_{\rm OUT}v_{\rm OUT}^2$~\citep[see, for example,][]{Gaspari:2017_uni}. From the kinematics of the multi-phase outer ring, we assume $v_{\rm OUT}= 2000$\kms\ and estimate $P_{\rm OUT}\sim 6\times 10^{42}$~erg s$^{-1}$. The outflowing power is compatible with the power of the last nuclear activity of \forn\, defined as $L_{\rm mech,jets}+L_{\rm rad}$. The mechanical power of the radio jets ($L_{\rm mech,jets}\sim 2.4\times 10^{42}$ erg\,s$^{-1}$) is inferred from their 1.44~GHz luminosity~\citep[$L_{\rm 1.44GHz,jets}$;][]{Cavagnolo:2010}, while the radiative power ($L_{\rm rad} \sim 3.4\times 10^{42}$erg s$^{-1}$) follows from the total luminosity of the \OIIIfs\ line of the outflow ($L_{\rm [OIII]5007}\gtrsim 9.8 \times 10^{38}$erg s$^{-1}$)~\citep[][]{Heckman:2004}.

The outflow power is compatible with the mechanical plus radiated power emitted by \forn\ in the last phase of activity ($L = L_{\rm mech} +L_{\rm rad} \gtrsim 6.0 \times 10^{42}$erg s$^{-1}$). This further indicates that the nuclear activity that generated the jets of \forn\ also entrained the gaseous outflow out to distances of $\sim 6$ kpc.

\subsection{The self-regulated flickering activity of \forn}

The radio continuum properties of \forn\ suggest that its nuclear activity is rapidly flickering. In the last $\sim$\,3 Myr two different episodes occurred, one that formed the jets and one that may still be ongoing~\citep{Maccagni:2020}.

In the previous Sections, we showed that, in the innermost 6 kpc of \forn, the gas experiencing condensation and turbulently raining co-exists with a regular ring which makes up most of the gas mass and gas outflowing as a consequence of the nuclear activity that occurred $\sim$\,3 Myr ago. The properties of both the outflow and the in-flowing gas are consistent with the predictions of the CCA model. Here, we argue that the rapid flickering of \forn\ is also self-regulated via CCA.

Within the framework of CCA, the macro-scale properties of the outflow can be linked to the mass that was accreted to fuel the nuclear activity. Feedback outbursts, propagating from the micro scale near the SMBH to the macro scales of the host galaxy/cluster, balance the cooling flow occurring at tens of kpc. Hence, AGN feedback must be significantly `energy conserving’ over large distances and long periods, with the power emitted at the micro scale ($P_{\rm out}$) roughly comparable to the power of the macro outflow ($P_{\rm OUT}$), and proportional to the luminosity of the X-ray halo. Following \citet{Gaspari:2017_uni}', the large scale outflow power is linked to the CCA mass accreted by the SMBH ($\dot{M}_{\rm cool}$) such as

\begin{equation}
\begin{split}
\label{eq:GS17}
	P_{\rm OUT} &= (1/2)\,\dot{M}_{\rm OUT}\,v_{\rm OUT}^2 \sim P_{\rm out}, \\
    P_{\rm out} &= (1/2)\,\dot{M}_{\rm out}\,v_{\rm out}^2 = \epsilon_{\rm BH}\,\dot{M}_{\rm cool}\,c^2,
\end{split}
\end{equation}

\noindent where $\epsilon_{\rm BH} = 10^{-3}\,(T_{\rm x}/2.5\,\rm keV)$ is the macro-scale mechanical efficiency, which is tied to the temperature of the X-ray emitting halo ($\sim$\,0.7 keV). Consequently the accretion rate (quenched cooling flow) needed to sustain the formation of the jets and generate the outflow is $\dot{M}_{\rm cool}\sim 0.4$~\msunyr.
Assuming as upper limit to the accretion the age of the jets ($\sim$\,3 Myr), the total accreted mass is of the order of $10^6$\,\msun, which is a fraction of the cold gas mass of the EW-stripe, as expected.
Based on Eq.~\ref{eq:GS17} and \citet{Gaspari:2017_uni} outflow scalings we can also estimate the properties of the micro feedback: $v_{\rm out} \approx 10^4$ \kms\ and $\dot M_{\rm out} \simeq 0.2$ \msunyr, which are consistent with the values of micro ultrafast X-ray outflows (\citealt{Tombesi:2013}).
From this, we can assess the entrainment factor $\eta = \dot M_{\rm OUT}/\dot M_{\rm out} \sim 25$, which indicates that the primary macro driver is the ionized/neutral component, rather than the molecular phase (with expected $\eta \gg 100$).
As shown by the above MUSE and MeerKAT $w80$ maps, such phases are indeed the main driver for the outflowing gas.

The k-plot, Taylor number and $C$-ratio provide major evidence that the EW-stripe is likely experiencing CCA. In these regions, according to CCA modelling, gas becomes  thermally unstable in a nonlinear way, generating substantial CCA rain unhindered by rotation, which will feed the SMBH via inelastic collisions as the clouds fall within $r < 300$ pc. Since the EW-stripe extends down to the innermost regions of \forn\ (Sect.~\ref{sec:feeding}) and is still condensing, it is reasonable to think that part of it recently fuelled the AGN. As CCA typically occurs not in a single episode of accretion, but through recursive precipitation episodes of low-mass feeding, this could explain the rapid flickering (a few Myr) of \forn\ in the low efficiency accretion regime (specifically, CCA typically induces pink-noise $1/f$ frequency power spectra). Overall, the several independent evidences at the macro and micro spatial scales, tied also to different epochs, have allowed us to achieve a full picture of the self-regulated AGN feeding and feedback cycle in \forn, with recursive CCA rainfalls and triggered outflows/jets.

\section{Summary and future prospects}
\label{sec:conclusions}

In this paper we have shown the distribution of the multi-phase ISM (neutral, molecular and ionised gas) in \forn, focusing on the innermost arcminute ($r\lesssim 6$ kpc). We studied the kinematics of the ISM based on spectral observations of the \HI\ from \meer, of the \couno\ from ALMA and of the optical emission lines (\hbeta, \OIIIqn, \OIIIfs, \NIIscq, \halpha, \NIIsco\ and \SIIdoublet) from MUSE. We used different methods that allowed us to identify both multi-phase gaseous inflows and outflows that are likely involved in the feeding and feedback processes of the AGN. 

In the innermost arcminute, most gas is distributed in a rotating nearly-edge on disk ($r\lesssim 4$ kpc). At outer radii ($4.5\lesssim r\lesssim 6.5$), the gas is found in a ring at the edge of the radio jets. The detailed analysis of the kinematics of the gas (Sect.~\ref{sec:kinIon}) suggests that the inner disk is rotating, while the outer ring is outflowing, ejected by the nuclear episode that also formed the jets. The cool phase is the most massive of the outflow ($M_{\rm out}\sim 2\times10^7$\msun). The macro-outflowing power ($1/2\dot{M_{\rm OUT}}v_{\rm OUT}^2$) is compatible with the power of the last phase of activity of \forn. The outflowing gas of the ring and along the jets is the result of the mass entrained by the winds ejected by the AGN. 

Besides the outflow, the multi-phase gas kinematics indicate that also other components are non-rotating,~\ie\ some filaments within the inner disk and the EW-stripe. The kinematical-plot (Sect.~\ref{sec:kIon}) for the multi-phase gas \citep[introduced by][]{Gaspari:2018} is a powerful method to identify all-at-once gaseous components that deviate from regular rotation. This allowed us to identify that the EW-stripe and filaments have narrow line-widths and line-shifts typical of gas undergoing CCA~(Sect.~\ref{sec:feeding}) and may be feeding the AGN. The $C$-ratio (Fig.~\ref{fig:cRatio}) indicates that in these regions the eddy-turnover timescale is roughly comparable to the cooling time and that condensation may efficiently occur, driving the gas to rain towards the centre and fuel the AGN. The mass of the in-falling filaments (dominated by the cold gas) is sufficient to feed the low-efficiency accretion regime of the current activity.

In \forn\ events of feeding and feedback are detected in the multi-phase gas over different scales (from $\lesssim 1$ kpc to $\sim 6$ kpc) and appear to co-exist in space and time, showing clear indications of preceding/ongoing and forthcoming condensation events. The connection between the properties of the macro-scale outflow and the in-falling gas suggests that the rapid flickering of the nuclear activity (1-3 Myr) is self-regulated by recurring CCA (\citealt{Gaspari:2017_cca}). Deeper X-ray observations may provide further information on this cycle by detecting the blueshifted absorption lines related to the micro AGN outflows. 

The kinematical analysis of the multi-phase ISM in \forn\ reveals both feeding and feedback events, and provides several quantitative evidences that CCA may be regulating its recurrent activity -- being arguably one of the most in-depth studies probing this mechanism. 
A complete study of the physical properties of the multi-phase gas, for example through the analysis of the ionised gas emission line ratios and by the high resolution observations of multiple transitions of the \CO, may allow us to gain further insights on the role of turbulence (and CCA) in sustaining the rapid AGN flickering and the effects of the outflow on the evolution of radio AGNs.

\begin{acknowledgements}
The authors thank the anonymous referee for the useful comments and suggestions. This project has received funding from the European Research Council (ERC) under the European Union’s Horizon 2020 research and innovation programme (grant agreement no.~679627). We are grateful to the full MeerKAT team at SARAO for their work on building and commissioning MeerKAT. The MeerKAT telescope is operated by the South African Radio Astronomy Observatory, which is a facility of the National Research Foundation, an agency of the Department of Science and Innovation. This paper makes use of the following ALMA data: ADS/JAO.ALMA\#2017.1.00129.S. ALMA is a partnership of ESO (representing its member states). NSF (USA) and NINS (Japan), together with NRC (Canada), MOST and ASAIA (Taiwan), and KASI (Republic of Korea), in cooperation with the Republic of Chile. The Joint ALMA Observatory is operated by ESO, AUI/NRAO and NAOJ. The VLA images at 4.8 and 14.4 GHz have been produced as part of the NRAO VLA Archive Survey, (c) AUI/NRAO. The National Radio Astronomy Observatory (NRAO) is a facility of the National Science Foundation, operated under cooperative agreement by Associated Universities, Inc. This research made use of Montage. It is funded by the National Science Foundation under Grant Number ACI-1440620, and was previously funded by the National Aeronautics and Space Administration's Earth Science Technology Office, Computation Technologies Project, under Cooperative Agreement Number NCC5-626 between NASA and the California Institute of Technology. This research was supported by the Munich Institute for Astro- and Particle Physics (MIAPP) of the DFG cluster of excellence `Origin and Structure of the Universe'. M.G. acknowledges partial support by NASA Chandra GO8-19104X/GO9-20114X and HST GO-15890.020-A grants. K.M. acknowledges JSPS KAKENHI Grant Numbers of 19J40004 and 19H01931. O.S.' research is supported by the South African Research Chairs Initiative of the Department of Science and Technology and National Research Foundation. 
\end{acknowledgements}

%
%


\bibliographystyle{aa} 
\bibliography{faGas.bib} 
\onecolumn

\begin{appendix} 

\section{Optical spectra and best fits of the emission lines of ionised gas}
\label{app:specMUSE}

{\centering
\null\vfill

	\includegraphics[trim = 0 0 0 0, height=23cm, width=16cm]{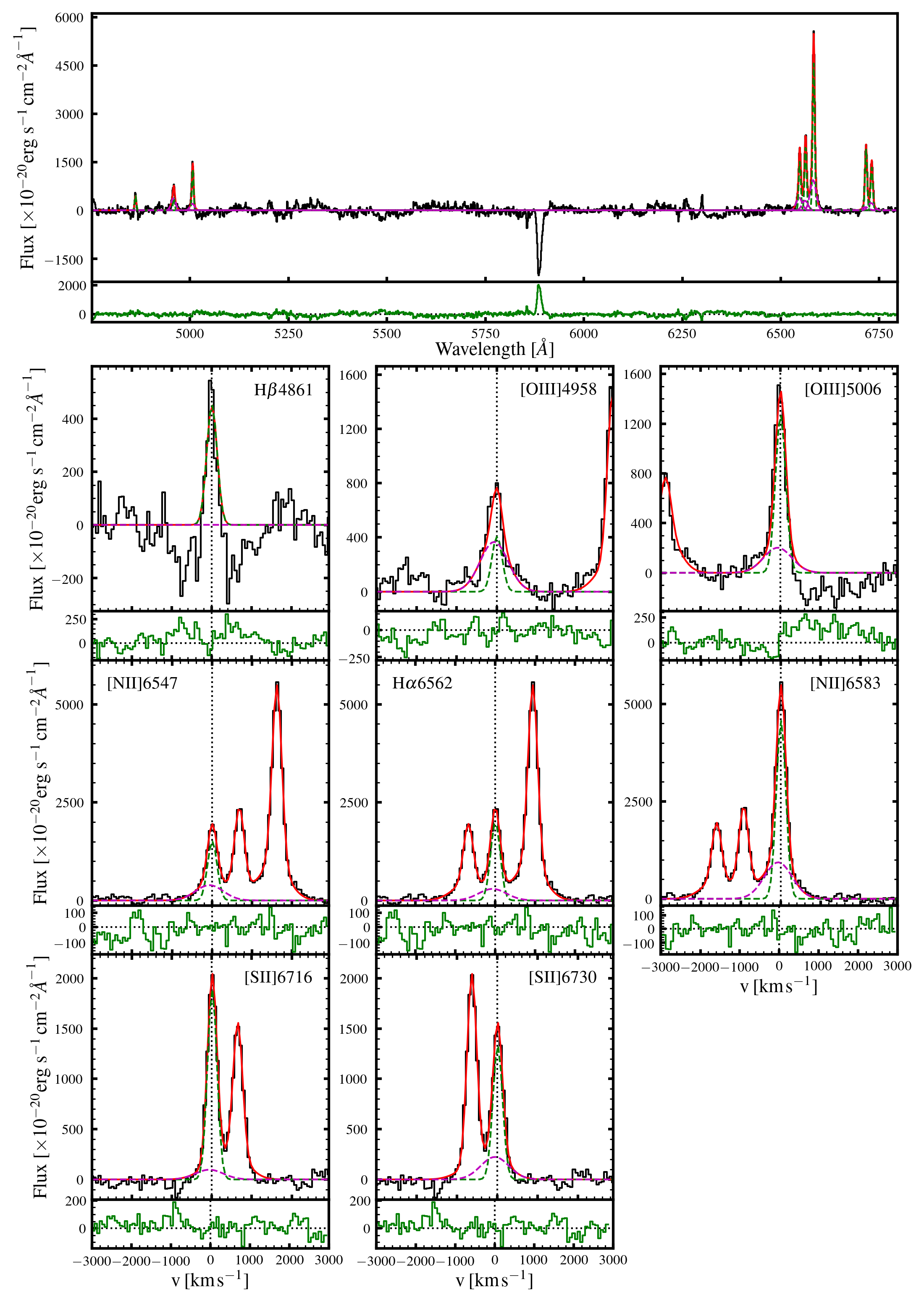}
\qquad
	\captionof{figure}{Centre of \forn.}
	\label{fig:specCtr}%

\vfill
	\includegraphics[trim = 0 0 0 0, height=24cm, width=16cm]{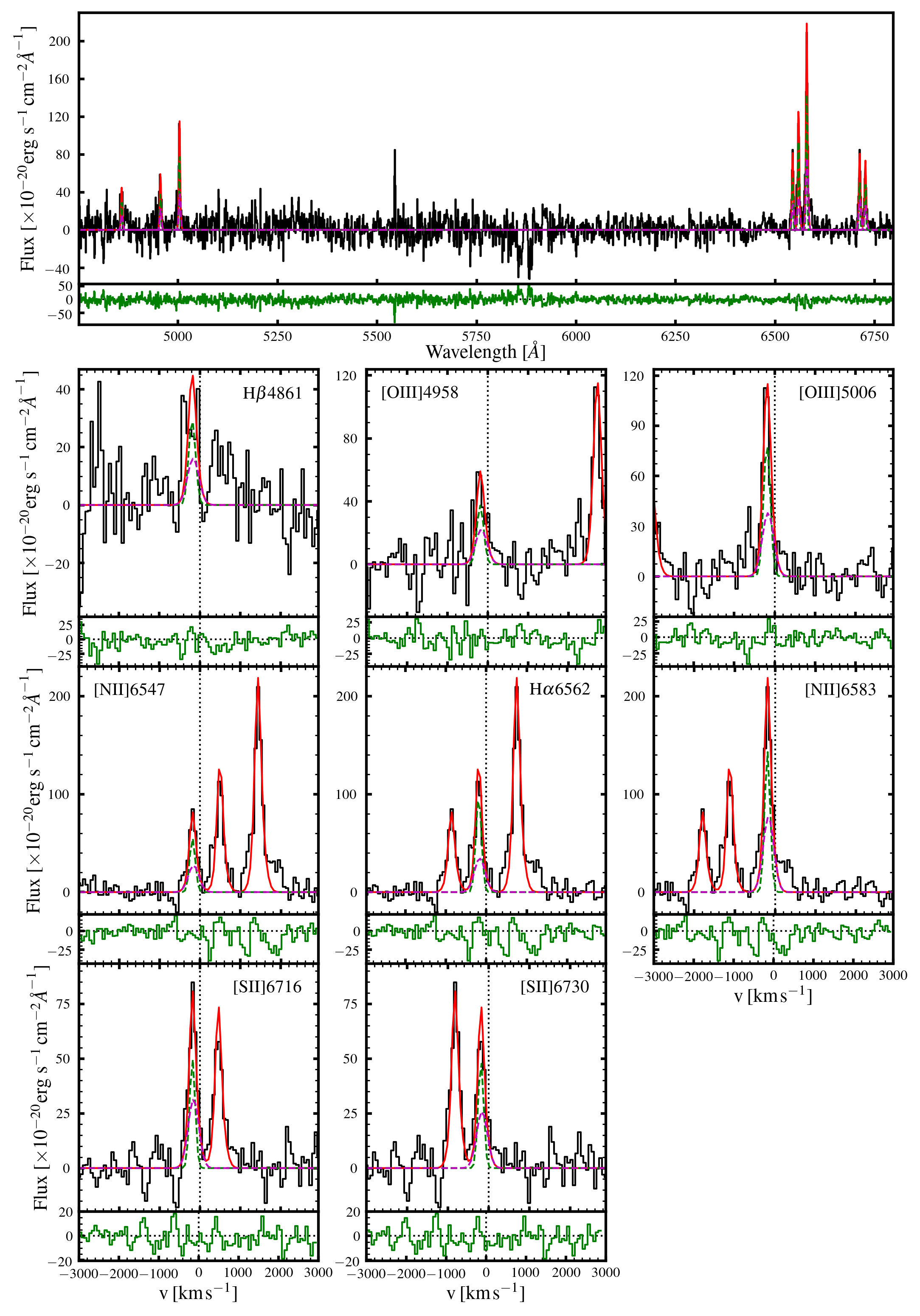}
	\captionof{figure}{Region with second Gaussian component along line A in the top left panel of Fig.~\ref{fig:momsIon}.}
	\label{fig:specG2}%
\vfill
	\includegraphics[trim = 0 0 0 0,height=24cm, width=16cm]{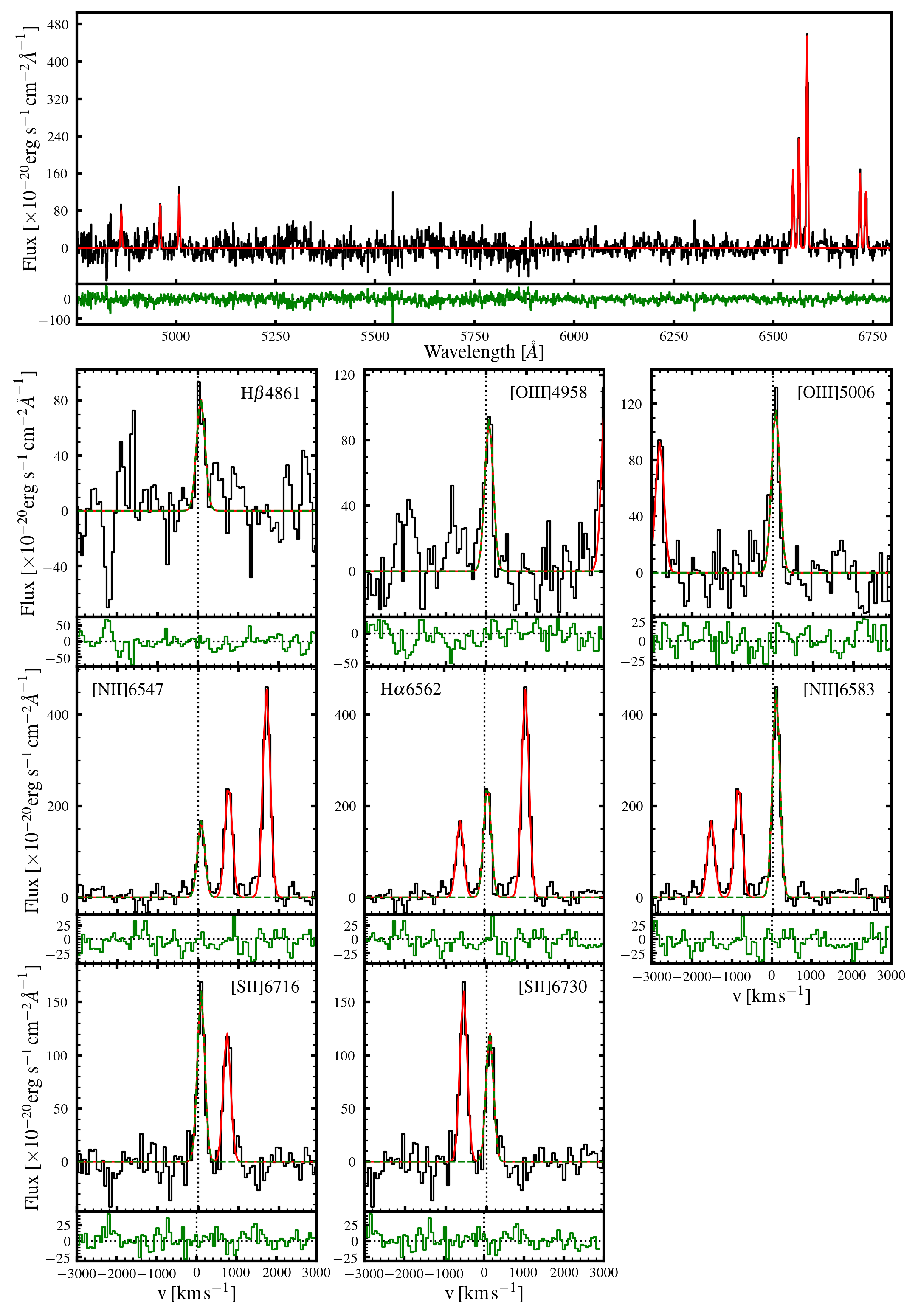}
	\captionof{figure}{Stripe, along the East of line B in the top left panel of Fig.~\ref{fig:momsIon}.}
	\label{fig:specStripe}%
\vfill
	\includegraphics[trim = 0 0 0 0, height=24cm, width=16cm]{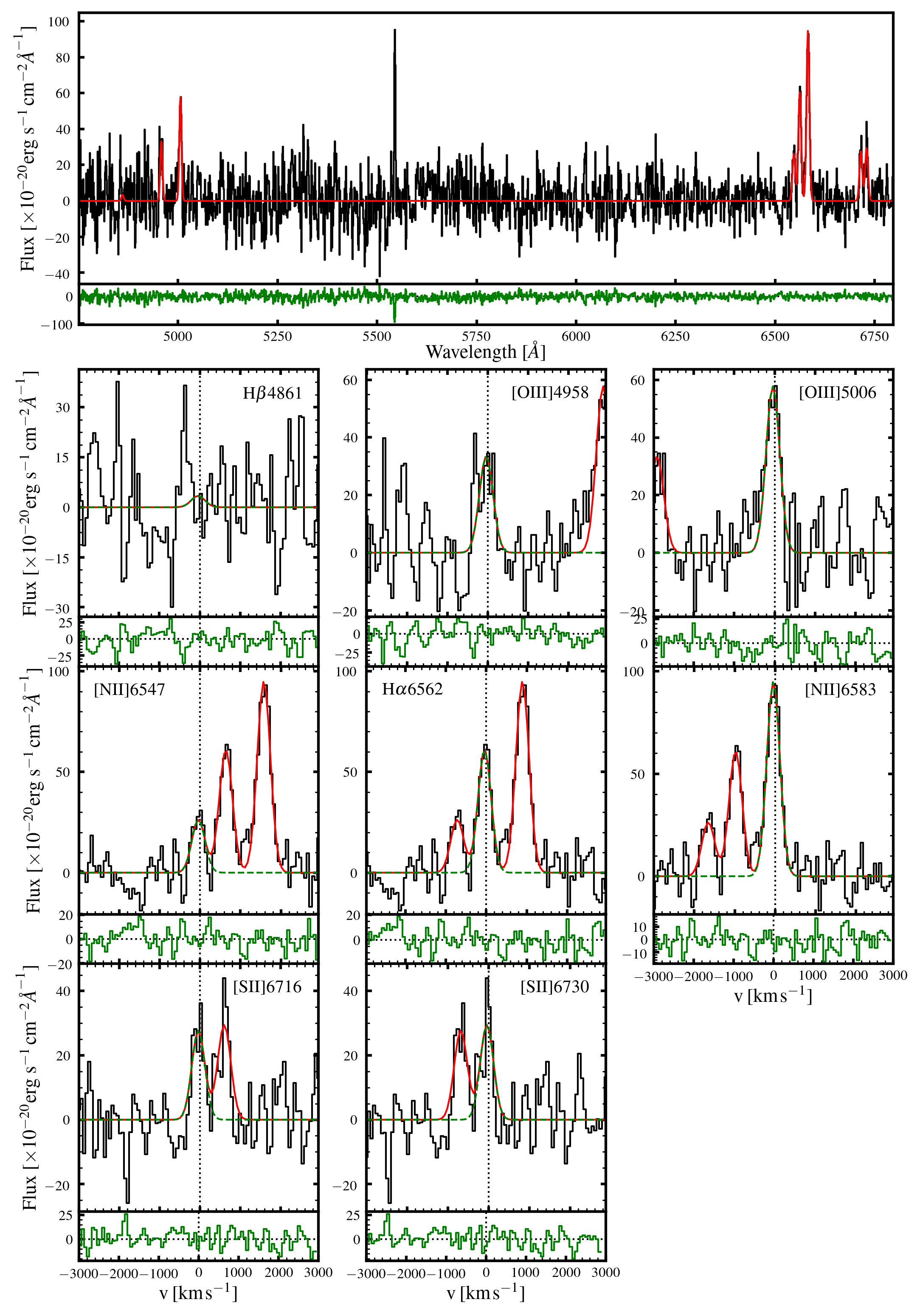}
	\captionof{figure}{Cloud, region C in the top left panel of Fig.~\ref{fig:momsIon}.}
	\label{fig:specCloud}%
\vfill
	\includegraphics[trim = 0 0 0 0,width=\textwidth]{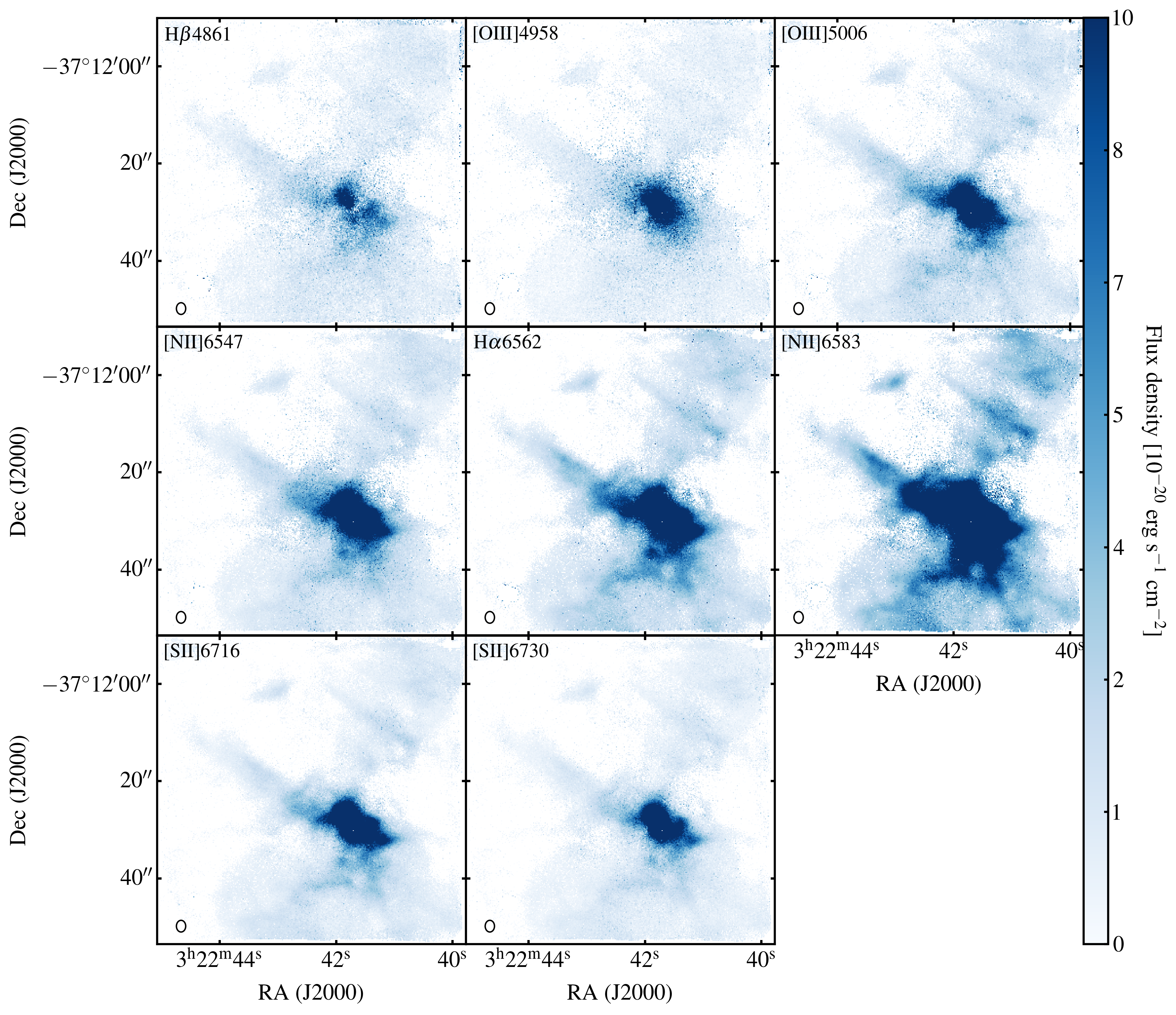}
	\captionof{figure}{Total flux density distribution of the fitted lines, on the same colourscale. In all lines the filamentary structure of the NS disk and EW-stripe can be appreciated. The highest emission is found in the centre and along the EW streams. The PSF of the MUSE observations ($2\arcsec$) is shown in the bottom left corner of each panel. Near the SE corner, the gaseous supernova remnant studied by~\citet{LopezCoba:2020} is visible in the in some ionised lines. The study of this, and other SNe remnants was the main science goal of the centre-field MUSE observations.}
	\vfill
	\vfill
	}
 \pagebreak
\section{Kinematical model of the ionised gas kinematics}
\label{app:modKin}

\begin{figure*}[tbh]
	\centering
	\includegraphics[trim = 0 0 0 0,width=\textwidth]{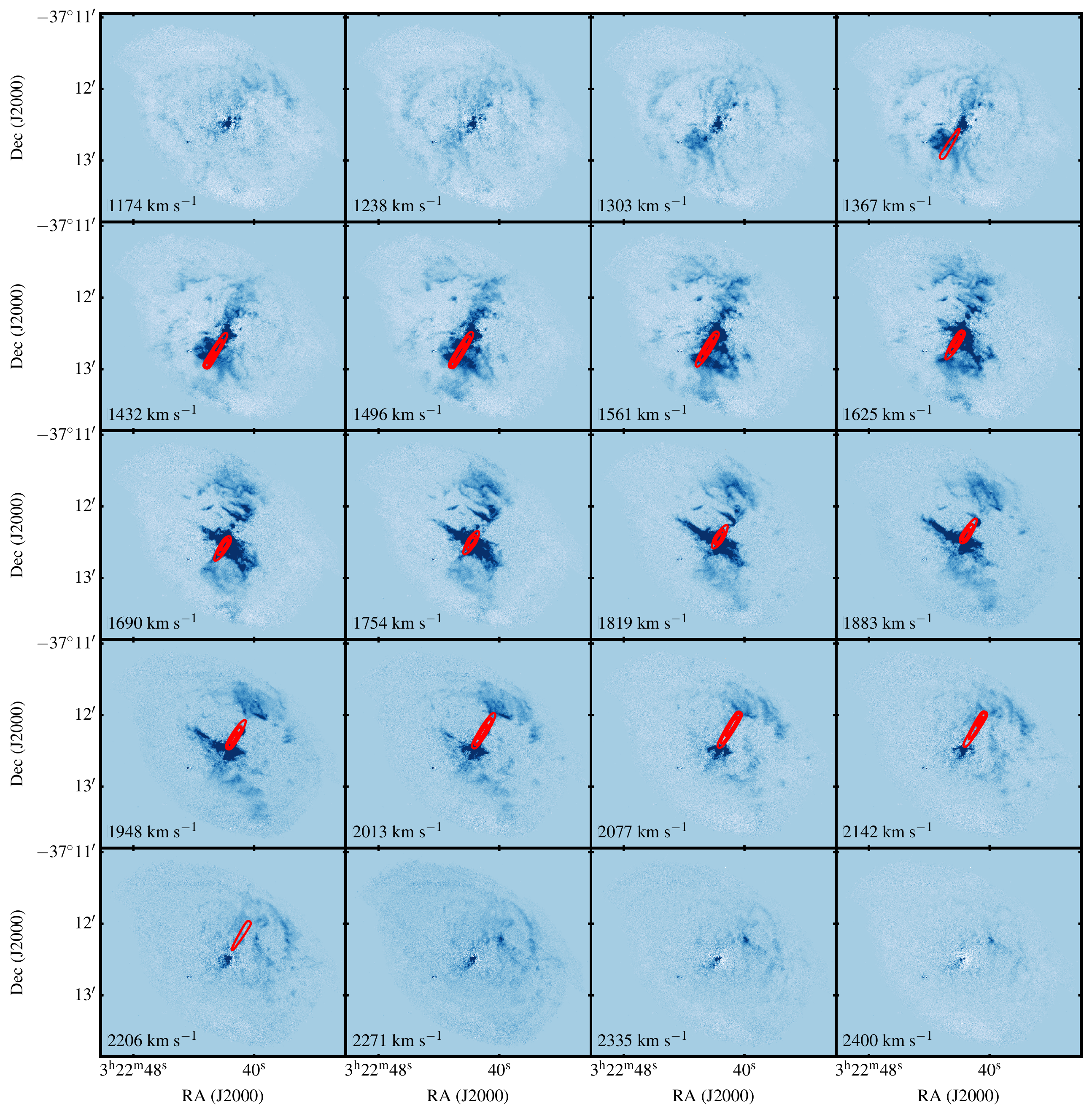}
	\caption{Consecutive channel maps extracted from the MUSE wide-field datacube. Overlaid in red are shown the contours of the tilted ring model that best reproduces the \NIIsco\ emission (as well as the \CO\ and \HI\ seen at lower resolution). Refer to Sect.~\ref{sec:modISM} for further details}
	\label{fig:chanMapsIonMod}%
\end{figure*}

\end{appendix}
\end{document}